\documentclass[useAMS,usenatbib]{mn2e}
\usepackage{graphicx}
\usepackage{color}
\usepackage{soul}
\usepackage{bm}
\usepackage{natbib}
\usepackage{amsmath}	
\usepackage{amssymb}	

\newcommand{\bec}[1]{\mbox{\boldmath $ #1$}}
\newcommand{\meanrho}{\overline{\rho}}
\newcommand{\meanBB}{\overline{\mbox{\boldmath $B$}}{}}{}
{}
\newcommand{\meanUU}{\overline{\mbox{\boldmath $U$}}{}}{}

\newcommand{\meanWW}{\overline{\mbox{\boldmath $W$}}{}}{}
\newcommand{\meanOmega}{\overline{\mbox{\boldmath $\Omega$}}{}}{}
\newcommand{\meanA}{\overline{A}}
\newcommand{\meanB}{\overline{B}}

\newcommand{\meanT}{\overline{T}}

\newcommand{\meanW}{\overline{W}}
\newcommand{\meanU}{\overline{U}}

\title[Alpha effect in stratified turbulence with large-scale shear]{Alpha effect
and dynamo in density-stratified turbulence with large-scale shear:
applications to protoplanetary discs and astrophysical clouds}

\author[I. Rogachevskii, N. Kleeorin]
 {I. Rogachevskii, N. Kleeorin,
 \\
 \\
Department of Mechanical Engineering,
Ben-Gurion University of Negev, POB 653, Beer-Sheva  8410530, Israel}

\begin{document}


\maketitle


\begin{abstract}
A joint effect of the density-stratified turbulence (or inhomogeneous turbulence) and
a large-scale shear for arbitrary Mach numbers results in
the $\alpha$ effect and mean-field dynamo action.
These effects also produce the effective  pumping velocity
of a large-scale magnetic field.
Compressibility of the turbulent velocity field (i.e., finite Mach number effect)
does not affect the contributions  to the $\alpha$ tensor caused by the
joint effect of inhomogeneity of turbulence
and a large-scale shear, but it influences the effective  pumping velocity of the mean magnetic field.
The isotropic part of the ${\bm \alpha}$ tensor is independent of the exponent
of the turbulent kinetic energy spectrum, while
its anisotropic part depends on this exponent.
This anisotropic part of the ${\bm \alpha}$ tensor depends on the latitudinal profile of
the large-scale shear velocity (differential rotation),
which may be important for dynamo operation in the upper parts of the solar and stellar convection zones.
There is also an additional contribution to the effective pumping velocity of the mean magnetic field
that is proportional to the product of the fluid density gradient and the divergence of the mean fluid velocity
caused, e.g., by collapsing (or expanding) astrophysical clouds.
Applications of these effects to protoplanetary discs, protogalactic
and protostellar clouds are discussed.
\end{abstract}


\maketitle


\section{Introduction}
\label{sect-1}

Large-scale magnetic fields in astrophysical turbulence can be generated
by combined effect of kinetic helicity and non-uniform (differential)
rotation or large-scale shear flow
\citep[see, e.g.,][]{Moffatt(1978),Parker(1979),Krause(1980),Zeldovich(1983),Ruzmaikin(1988),Ruediger(2013),MD2019,RI21,SS21}.
The kinetic helicity in turbulence can be produced in rotating inhomogeneous
or density stratified turbulence.
The alternative to rotation  is the large-scale shear which is
in combination with the density-stratified turbulence
(or inhomogeneous turbulence) can produce
the kinetic helicity and the $\alpha$ effect.

Examples of astrophysical systems where large-scale shear motions
place an important role are
protoplanetary discs  \citep[see, e.g.,][]{HB98,EKR98,PPK11,HUB16,HOP16,HOP16b,KR25},
colliding protogalactic clouds and merging protostellar clouds
\citep[see, e.g.,][]{CH91,CH93,WBL98,BWL02,RKCL06},
as well as solar and stellar convective zones \citep[see, e.g.,][]{Parker(1979),Krause(1980)}.
In such systems large-scale shear motions coexist with small-scale turbulence.
Interaction between large-scale shearing motions and density stratified or inhomogeneous turbulence
causes a non-zero $\alpha$ effect and generation of large-scale magnetic field.
In addition to the large-scale shear motions, there can be collapsing
or expanding astrophysical clouds or disks.
Typical examples of such astrophysical systems are gravitational collapse of young stars,
expanding Universe, and supernova explosions resulting to production of turbulence
in galaxies and formation of expanding astrophysical clouds.

The $\alpha$ effect and effective pumping velocity in {\it inhomogeneous and incompressible turbulence with a large-scale shear}
have been determined by \cite{RK03} using the spectral $\tau$ approach for large fluid and magnetic Reynolds numbers.
These effects have been also studied applying the quasi-linear approach (or the second-order correlation approximation) by \cite{RS06}.
This approach is valid for small fluid and magnetic Reynolds numbers
or for high conductivity limit and small Strouhal numbers.

In addition, a mean-field theory for a pumping effect of the mean magnetic field
in {\it helical homogeneous turbulence with large-scale shear}
has been also developed by \cite{RKK11}, applying various analytical methods.
In particular, they have used the quasi-linear approach, the path-integral technique, and the spectral $\tau$
approach, and have found that the effective pumping velocity is proportional to the product of $\alpha$ effect and large-scale vorticity
associated with the large-scale shear.
Direct numerical simulations of helical turbulence with large-scale shear in different ranges of hydrodynamic and
magnetic Reynolds numbers have found the effective pumping velocity of the mean magnetic field
by a kinematic test-field method in agreement with the theoretical predictions by \cite{RKK11}.

However, the $\alpha$ effect and effective pumping velocity have not yet been derived
for {\it a density-stratified non-helical
background} turbulence with large-scale shear and for arbitrary Mach numbers.
In the present study, we investigate these effects applying the spectral $\tau$ approach.
We find that the isotropic part of the ${\bm \alpha}$ tensor
is independent of the exponent of the turbulent kinetic energy spectrum.
The joint effect of density-stratified turbulence and large-scale shear
 also produce the effective  pumping velocity
of a large-scale magnetic field.

There are also additional contributions to the effective pumping velocity
${\bm V}^{\rm eff} \propto {\bm \lambda} \,  {\rm div} \meanUU$
in density stratified turbulence, or ${\bm V}^{\rm eff} \propto {\bm \Lambda} \,  {\rm div} \meanUU$
in inhomogeneous turbulence, which can arise in collapsing (or expanding) astrophysical turbulent clouds.
Here ${\bm \lambda} = - {\bm \nabla} \ln \meanrho$ describes the fluid density
stratification,
${\bm \Lambda} = {\bm \nabla} \ln \langle {\bm u}^2 \rangle^{(0)}$ describes
inhomogeneous background turbulence,
where the angular brackets imply an ensemble averaging,
$\meanrho$ is the mean fluid density and $\meanUU$ is the mean velocity.
Note that magnetic field amplification during a turbulent collapse have been recently
studied  by \cite{BN25} and \cite{IBSS25}
(see also references therein).
The ${\bm \alpha}$ tensor is independent of ${\rm div} \meanUU$, i.e.,
it is independent of the effects of collapsing or expanding of clouds.
We apply the obtained results related to the ${\bm \alpha}$ tensor and effective pumping velocity
of the large-scale magnetic field to protoplanetary discs, colliding protogalactic clouds and
merging protostellar clouds.

This paper is organized as follows.
In Sec.~\ref{sect-2} we discuss the assumptions and the procedure for the derivation
of the turbulent electromotive force (EMF).
In Sec.~\ref{sect-3} we determine the $\alpha$ effect and the effective pumping velocity of the large-scale magnetic field
in a density-stratified turbulence with a large-scale shear flow.
For comparison, in Sec. ~\ref{sect-4} we find the $\alpha$ effect and the effective pumping
velocity in an inhomogeneous turbulence with a large-scale shear flow.
In Sec. ~\ref{sect-5} we consider compressible turbulence with large-scale shear and
calculate the $\alpha$ effect and the effective pumping
velocity for arbitrary Mach numbers.
In Sec. ~\ref{sect-6} we discuss applications of the obtained results to protoplanetary discs
and astrophysical clouds.
Finally,  in Sec.~\ref{sect-7} we draw the concluding remarks.
In Appendix~\ref{Appendix A} we give identities used for derivation of the $\alpha$
effect and the effective pumping velocity of the mean magnetic field.

\section{Governing equations and method of derivations}
\label{sect-2}

To determine the $\alpha$ effect and effective pumping velocity of
the mean magnetic field in a density-stratified, inhomogeneous and compressible turbulence
with non-uniform large-scale flow,
we follow the approach described by \cite{RK04} and \cite{KR22} [see also book by \cite{RI21}].
We consider turbulence with
large fluid and magnetic Reynolds numbers, so that
the Strouhal number (the ratio of the characteristic turbulent time $\tau_0$
to the turn-over time $\ell_0/u_0$) is of the order of unity.
Here $\ell_0$ and $u_0$ are the integral turbulence scale
and characteristic turbulent velocity at the integral scale.

We apply the mean-field approach, i.e.,
we assume that there is a separation of spatial and temporal scales,
$\ell_0 \ll L_B$ and $\tau_0 \ll t_B$, where
$L_B$ and $t_B$ are the spatial and temporal scales
characterising the variations of the mean magnetic field.
We also use the multi-scale approach \citep{RS75}.
In the framework of the mean-field approach, we separate
magnetic field ${\bm B} = \meanBB + {\bm b}$ and velocity field
${\bm U} =\meanUU + {\bm u}$ into mean field and fluctuations,
where $\meanBB = \langle{\bm B} \rangle $ is the mean magnetic field,
$\meanUU = \langle {\bm U} \rangle$ is the mean fluid velocity,
${\bm b}$ and ${\bm u}$ are magnetic and velocity fluctuations, respectively.
In similar fashion, we separate fluid density and pressure.
Here we use the Reynolds averaging, which implies that $\langle {\bm u} \rangle=0$,
$\langle {\bm b} \rangle=0$, etc.

We determine contributions to
the turbulent electromotive force $\langle {\bm u} \times {\bm b}\rangle$
caused by the non-uniform large-scale flow $\meanUU $.
To this end, we use the momentum equation for velocity fluctuations
${\bm u}$ and the induction equation for magnetic fluctuations ${\bm b}$ as
\begin{eqnarray}
&& {\partial {\bm u} \over \partial t} = - (\meanUU \cdot {\bm \nabla}) {\bm u} - ({\bm u} \cdot
{\bm \nabla}) \meanUU - {{\bm \nabla} p_{\rm tot} \over \meanrho}  + {\bm F}_\nu + {\bm F}
\nonumber \\
&& \quad + {1 \over \mu_0 \meanrho} \Big[({\bm b} \cdot
{\bm \nabla}) \meanBB
+ (\meanBB \cdot {\bm \nabla}){\bm b}\Big]
  + {\bm u}^{\rm (N)} ,
\label{B0}
\end{eqnarray}

\begin{eqnarray}
&& {\partial {\bm b} \over \partial t} = {\bm \nabla} \times
\Big[\meanUU \times {\bm b} + {\bm u} \times \meanBB
- \eta {\bm \nabla} \times {\bm b}\Big] + {\bm b}^{\rm (N)},
\label{B1}
\end{eqnarray}
where $\eta$ is the magnetic diffusion due to electrical conductivity of fluid,
$\meanrho $ is the mean fluid density, $\mu_0$ is the magnetic permeability of the fluid,
${\bm F} $ is a random external stirring force, $ p_{\rm tot} = p + \mu_0^{-1} \,(\meanBB
\cdot {\bm b}) $ are fluctuations of the total pressure, $p$ are
fluctuations of the fluid pressure, ${\bm u}^{\rm (N)}$ and
${\bm b}^{\rm (N)}$ are the nonlinear terms. The velocity ${\bm u}$
satisfies to the continuity equation.
Generally, all mean quantities depend on coordinate and time.

Equation~(\ref{B0}) is written for the case when fluctuations of the fluid density
are much smaller in comparison with the mean fluid density.
For simplicity, the mean fluid velocity in this study describes only two effects:
(i) the imposed large-scale shear
and (ii) the effects of collapsing or expanding of clouds
which can be described by large-scale motions with
a non-zero ${\rm div} \meanUU$ (see  section~\ref{sect-5}).

Using equations~(\ref{B0})--(\ref{B1}), we derive equations for the cross-helicity tensor
$g_{ij}({\bm k}) = \langle u_i(t, {\bm k}) \, b_j(t, -{\bm k}) \rangle$ and the tensor
$f_{ij}({\bm k}) = \langle u_i(t, {\bm k}) \, u_j(t, -{\bm k}) \rangle$
for the velocity fluctuations in a Fourier space.
Since our goal is to derive only expressions for the $\alpha$ effect and the effective pumping velocity
of the mean magnetic field, we neglect terms proportional to spatial derivatives
of the mean magnetic field in these equations.
We consider the kinematic dynamo problem and do not discuss the nonlinear effects, so we do not need
evolutionary equation for the  tensor for magnetic fluctuations
$b_{ij}({\bm k}) = \langle b_i(t, {\bm k}) \, b_j(t, -{\bm k}) \rangle$.

Equations for the second-order moments include the first-order spatial
differential operators applied to the third-order
moments $\hat{\cal M} g_{ij}^{(III)}({\bm k})$ and $\hat{\cal M} f_{ij}^{(III)}({\bm k})$
appearing due to the nonlinear terms.
Therefore, a problem arises how to close the system, i.e.,
how to express the third-order moments through the second-order moments, $g_{ij}$ and  $f_{ij}$
denoted as $F^{(II)}$. We use the spectral $\tau$ approach
\citep{PFL76,O70,RI21},
which postulates that the deviations of the third-order moments, denoted as $\hat{\cal M} F^{(III)}({\bm k})$, from the contributions to these terms afforded by a background turbulence, $\hat{\cal M} F^{(III,0)}({\bm k})$, can be expressed through the similar deviations of the second moments, $F^{(II)}({\bm k}) - F^{(II,0)}({\bm k})$ as
\begin{eqnarray}
&& \hat{\cal M} F^{(III)}({\bm k}) - \hat{\cal M} F^{(III,0)}({\bm
k}) = - {1 \over \tau_r(k)} \, \Big[F^{(II)}({\bm k})
\nonumber\\
&& \quad - F^{(II,0)}({\bm k})\Big] \,,
\label{B2}
\end{eqnarray}
where $\tau_r(k)$ is the scale-dependent relaxation time, which can be identified with the turbulent time $\tau(k)$ of velocity fluctuations for large fluid and magnetic Reynolds numbers
\citep[see, e.g.,][]{RI21}.
The functions with the superscript $(0)$ correspond to the background turbulence with a zero mean magnetic field
and a zero large-scale shear.
The background turbulence is assumed to be stationary in statistical sense.

The turbulent time in the ${\bm k}$ space is defined as in the Kolmogorov-like
turbulence: $\tau(k) = 2\tau_0 \, \bar \tau(k)$ \citep[see, e.g.,][]{RI21},
where $\tau_0= \ell_0 /u_0$ is the characteristic turbulent time,
the function $\bar \tau(k) =(k / k_{0})^{1-q}$, and
the turbulent kinetic energy spectrum in the inertial range of wave numbers $k_0<k<k_\nu$ is
$E(k) = - d \bar \tau(k) / dk = (q-1)  \, k_0^{-1} \, (k / k_{0})^{-q}$.
Here the wave number $k_{0} = 1 / \ell_0$,
and $u_0 = \left[\left\langle {\bm u}^2 \right\rangle^{(0)}\right]^{1/2} \equiv u_{\rm rms}$,
the wave number $k_{\nu}=\ell_{\nu}^{-1}$,
the length $\ell_{\nu}$ is the Kolmogorov (viscous) scale.
The exponent $q=5/3$ corresponds to the Kolmogorov spectrum.
Generally, the exponent $q$ can be varied within the interval $1<q<3$.
The condition $q>1$ corresponds to finite kinetic energy
for very large fluid Reynolds numbers, while $q<3$ corresponds to finite
dissipation of the turbulent kinetic energy at the viscous scale.

Validation of the $\tau$ approaches applied in the ${\bm k}$ space (so called the spectral $\tau$ approximation) or in the physical space (so called the minimal $\tau$ approximation) for different situations has been performed in various numerical simulations
\citep{Brandenburg(2005),BS05b,BRK12b,BGKR13,BKR16,RKK11,RKB12,RRB17,RKB18,SRB18}.

When the mean magnetic field is zero, $g_{ij}^{(0)}({\bm k})=0$ and the turbulent electromotive force vanishes.
Consequently, equation~(\ref{B2}) reduces to $\hat{\cal M} g_{ij}^{(III)}({\bm k}) = - g_{ij}({\bm k}) / \tau(k)$ and $\hat{\cal M} f_{ij}^{(III)}({\bm k}) -\hat{\cal M} f_{ij}^{(III,0)}({\bm k})= - [f_{ij}({\bm k}) - f_{ij}^{(0)}({\bm k})]/ \tau(k)$.
We assume that the characteristic time of variation of the second-order moments $g_{ij}({\bm k})$ and $f_{ij}({\bm k})$
are substantially larger than the correlation time $\tau(k)$ for all turbulence scales.

Therefore, the contributions $g_{ij}^{(S)}({\bm  k})$ to the cross-helicity tensor $\langle u_i ({\bm  k}) \,  b_j(-{\bm k}) \rangle$
caused by turbulence with a non-zero large-scale shear are given by
\begin{eqnarray}
&& g_{ij}^{(S)}({\bm  k})  = \tau(k) \biggl\{ -{\rm  i} \, ({\bm k} \cdot \meanBB) \,   \biggl[\tau(k) \, J_{ijmn}(\meanUU)
\, \Big(f_{mn}^{(0)}({\bm k})
\nonumber\\
&&
- b_{mn}^{(0)}({\bm k}) \Big)  + f_{ij}^{(S)}({\bm k})
- b_{ij}^{(S)}({\bm k})  \biggr] + {1 \over 2} (\meanBB \cdot {\bm \nabla}) f_{ij}^{(S)}({\bm k})
\nonumber\\
&& + \overline{B}_j \biggl[i k_n - \lambda_n - {1 \over 2} \nabla_n \biggr]f_{in}^{(S)} \biggr\},
\label{B3}
\end{eqnarray}
where the tensor $f_{ij}^{(0)}({\bm k},{\bm R})=\left\langle u_i({\bm k}) \, u_j(-{\bm k}) \right\rangle^{(0)}$
describes velocity fluctuations in the background turbulence.
Here we also take into account small-scale dynamo that
generates magnetic fluctuations in the background turbulence with a zero mean magnetic field, and
characterised by the tensor $b_{ij}^{(0)}({\bm k})=\left\langle b_i({\bm k}) \, b_j(-{\bm k}) \right\rangle^{(0)}$.
The contributions $f_{ij}^{(S)}({\bm  k})$ and $b_{ij}^{(S)}({\bm k})$ of the large-scale shear on velocity
and magnetic fluctuations are given by
\begin{eqnarray}
f_{ij}^{(S)}({\bm  k})  = \tau(k) \, I_{ijmn}(\meanUU) \, f_{mn}^{(0)}({\bm  k}),
\label{B4}
\end{eqnarray}
\begin{eqnarray}
b_{ij}^{(S)}({\bm  k})  = \tau(k) \, E_{ijmn}(\meanUU) \, b_{mn}^{(0)}({\bm  k}) ,
\label{NNB4}
\end{eqnarray}
where the tensors $I_{ijmn}(\meanUU)$, $J_{ijmn}(\meanUU)$ and $E_{ijmn}(\meanUU)$ are given by
equations~(\ref{B5})--(\ref{NNB5})  in Appendix~\ref{Appendix A}.

In the next sections, using equations~(\ref{B3})--(\ref{NNB4}),
we determine the $\alpha$ effect and effective pumping velocity of
the mean magnetic field produced by a large-scale shear
imposed on various kinds of small-scale background turbulence:
\begin{itemize}
\item{a density-stratified turbulence (Sect.~\ref{sect-3});}
\item{inhomogeneous turbulence (Sect.~\ref{sect-4}) and}
\item{compressible density-stratified and inhomogeneous turbulence (Sect.~\ref{sect-5}).}
\end{itemize}

\section{Density-stratified turbulence with large-scale shear}
\label{sect-3}

In this section, we consider a density-stratified turbulence with large-scale shear
and low-Mach-number flows.
The velocity ${\bm u}$ satisfies to the continuity equation applied
in the anelastic approximation:
\begin{eqnarray}
{\bm \nabla} \cdot (\meanrho \, {\bm u}) = 0 ,
\label{AN1}
\end{eqnarray}
so that div ${\bm u} = {\bm \lambda} \cdot {\bm u}$, where ${\bm \lambda}
= - ({\bm \nabla} \meanrho)/\meanrho$.
We remind that we consider the case when fluctuations $\rho'$ of the fluid density
are much smaller in comparison with the mean fluid density ($|\rho'| \ll \meanrho$)
and $|\rho' \, \meanUU| \ll \meanrho \, |{\bm u'}|$.

We use the following model for the second moment, $f_{ij}^{(0)}({\bm k},{\bm R})=\left\langle u_i({\bm k}) \, u_j(-{\bm k}) \right\rangle^{(0)}$  of velocity fluctuations in a density-stratified
homogeneous background turbulence in a Fourier space:
\begin{eqnarray}
f_{ij}^{(0)} = {\left\langle {\bm u}^2 \right\rangle^{(0)} \, E(k) \over 8 \pi k^2} \, \biggl[\delta_{ij} - k_{ij}
+ {{\rm i} \over k^2} \, \big(\lambda_i k_j - \lambda_j k_i\big)\biggr]
\label{B7}
\end{eqnarray}
\citep[see, e.g.,][]{RI21},
where $\delta_{ij}$ is the Kronecker unit tensor and $k_{ij}=k_i \, k_j / k^2$.

We also take into account the small-scale dynamo in the background turbulence.
To this end, we use the following model for the second moment, $b_{ij}^{(0)}({\bm k})=\left\langle b_i({\bm k}) \, b_j(-{\bm k}) \right\rangle^{(0)}$  of magnetic fluctuations:
\begin{eqnarray}
b_{ij}^{(0)} = {\left\langle {\bm b}^2 \right\rangle^{(0)} \, E_{\rm M}(k) \over 8 \pi k^2} \, \Big(\delta_{ij} - k_{ij}\Big) ,
\label{MB7}
\end{eqnarray}
where  $E_{\rm M}(k)$ is the energy spectrum function of magnetic fluctuations.
For simplicity, we assume that $E_{\rm M}(k)=E(k)$ for $k \geq \ell_{\rm M}^{-1}$ and $E_{\rm M}(k)=0$
for $\ell_0^{-1}< k<\ell_{_{\rm M}}^{-1}$,
where $\ell_{_{\rm M}}$ is the characteristic scale of the localization of the maximum of magnetic energy caused by the small-scale dynamo.
Note that equation~(\ref{MB7}) for magnetic fluctuations of the background turbulence
does not contain terms proportional to ${\bm \lambda}$
since div ${\bm b} = 0$, while equation~(\ref{B7}) for velocity fluctuations includes terms
proportional to ${\bm \lambda}$ due to the continuity equation~(\ref{AN1}): div $(\meanrho \, {\bm u}) = 0$.

In the present study we only derive expressions for the $\alpha$ effect
and the effective pumping velocity, so we neglect terms proportional to spatial derivatives
of the mean magnetic field in the turbulent electromotive force.
We take into account that the terms in $g_{ij}^{(S)}({\bm k})$ with
symmetric tensors with respect to the indexes "i" and "j" do not
contribute to the turbulent electromotive force because ${\cal E}_{m} =
\varepsilon_{mij} \, \int g_{ij}^{(S)}({\bm k}) \,d{\bm k}$.

The contributions to the turbulent electromotive force
caused by density-stratified turbulence with large-scale shear
can be written as ${\cal E}_i^{(\lambda)}
= a_{ij}^{(\lambda)} \, \meanB_j$, where $a_{ij}^{(\lambda)}$ is given by equation~(\ref{MD1}) in Appendix~\ref{Appendix A}.
Now we determine the tensor $\alpha_{ij}^{(\lambda)} = (a_{ij}^{(\lambda)} + a_{ji}^{(\lambda)})/2 $ and the effective  pumping velocity $V^{\rm eff}_{n}({\bm \lambda}) = - \varepsilon_{ijn} \, a_{ij}^{(\lambda)} / 2$ of the mean magnetic field
in a density-stratified turbulence with a large-scale shear:
\begin{eqnarray}
&& \alpha_{ij}^{(\lambda)} = {\ell_0^2 \over 45} \, \biggl[13\left({\bm \lambda} \cdot \meanWW\right) \,  \delta_{ij}
-2 \Big(\lambda_i \, \meanW_j + \lambda_j \, \meanW_i\Big)
\nonumber\\
&& \quad +
(4 q -7) \, \lambda_m \Big(\varepsilon_{imn} \, (\partial \meanU)_{nj}
+ \varepsilon_{jmn} \, (\partial \meanU)_{ni}\Big) \biggr] ,
\label{B9}
\end{eqnarray}

\begin{eqnarray}
&& V^{\rm eff}_{i}({\bm \lambda}) ={\ell_0^2 \over 45} \, \biggl[ 5 \left({\bm \lambda}\times \meanWW \right)_{i}
+2(8q + 11)\, \lambda_m (\partial \meanU)_{mi}\biggr] ,
\nonumber\\
\label{B10}
\end{eqnarray}
where $\meanWW={\bm \nabla} \times \meanUU$ is the mean vorticity,
$(\partial \meanU)_{ij}=(\nabla_i \meanU_j + \nabla_j \meanU_i)/2$,
and $\varepsilon_{ijk}$ is the fully antisymmetric Levi-Civita tensor.
Here the gradient of the mean velocity $\nabla_i \meanU_j$ is decomposed into symmetric
$(\partial \meanU)_{ij}$ and
antisymmetric $\varepsilon_{ijp} \, \meanW_p/2$ parts, i.e.,
$\nabla_i \meanU_j=(\partial \meanU)_{ij} + \varepsilon_{ijp} \, \meanW_p/2$.
In the present study we consider only a weak large-scale shear ($\meanW \tau_0 \ll 1$),
and neglect the second-order derivatives of the mean velocity $\meanUU$.

Equations~(\ref{B9})--(\ref{B10}) are given in the absence of the small-scale dynamo.
The tensor $\alpha_{ij}^{(\lambda,b)}  = (a_{ij}^{\rm (M,\lambda)} + a_{ji}^{\rm (M,\lambda)})/2 $ and the effective  pumping velocity $V^{\rm eff}_{n}({\bm \lambda},b) = - \varepsilon_{ijn} \, a_{ij}^{\rm (M,\lambda)} / 2$
caused by the small-scale dynamo are given by
\begin{eqnarray}
&& \alpha_{ij}^{(\lambda,b)} = - {\ell_0^2 \over 45} \, \biggl({\ell_{_{\rm M}} \over \ell_0}\biggr)^{3(q-1)}
\biggl[{ \left\langle {\bm b}^2 \right\rangle^{(0)} \over \mu_0 \meanrho \, \left\langle {\bm u}^2 \right\rangle^{(0)} } \biggr]
\biggl[\left({\bm \lambda} \cdot \meanWW\right)  \delta_{ij}
\nonumber\\
&& \; + \lambda_i  \meanW_j + \lambda_j \meanW_i - \lambda_m \Big(\varepsilon_{imn} \, (\partial \meanU)_{nj}
+ \varepsilon_{jmn} \, (\partial \meanU)_{ni}\Big) \biggr] ,
\nonumber\\
\label{MB9}
\end{eqnarray}
\begin{eqnarray}
&& V^{\rm eff}_{i}({\bm \lambda},b) = {\ell_0^2 \over 45} \, \biggl({\ell_{_{\rm M}} \over \ell_0}\biggr)^{3(q-1)}
\biggl[{ \left\langle {\bm b}^2 \right\rangle^{(0)} \over \mu_0 \meanrho \, \left\langle {\bm u}^2 \right\rangle^{(0)} } \biggr] \,
\lambda_m (\partial \meanU)_{mi}  ,
\nonumber\\
\label{MB10}
\end{eqnarray}
where
$a_{ij}^{\rm (M,\lambda)}$ is given by equation~(\ref{BAD4}) in Appendix~\ref{Appendix A}.
As follows from equations~(\ref{MB9})--(\ref{MB10}), magnetic fluctuations caused by the small-scale dynamo
decrease the $\alpha$ effect.
However, the contributions due to the small-scale dynamo are smaller than those caused by the velocity fluctuations since
$\ell_{_{\rm M}} \ll \ell_0$ \citep{BRS23} and
$\left\langle {\bm b}^2 \right\rangle^{(0)}/\mu_0 \leq  \meanrho \, \left\langle {\bm u}^2 \right\rangle^{(0)}$.
Indeed, as follows from direct numerical simulations by \cite{BRS23}, the ratio of the wave numbers
$k_{_{\rm M}}/k_\nu \propto {\rm Pm}^{0.6}$
for Pm $\geq 2$ and $k_{_{\rm M}}/k_\nu \propto {\rm Pm}$
for Pm $ < 1$ [see Fig. 2 in \cite{BRS23}].
Here ${\rm Pm}$ is the magnetic Prandtl number, $\ell_{_{\rm M}} = k_{_{\rm M}}^{-1}$
and the wavenumber $k_\nu$ is based on the Kolmogorov scale.
However, $\ell_{_{\rm M}} \ll \ell_0$ [see Table 1 in \cite{BRS23}].

The kinetic $\alpha$  effect based on the isotropic part ($\propto \delta_{ij}$) of the alpha tensor  is given by
\begin{eqnarray}
\alpha^{(\lambda)} = {13 \over 45} \, \ell_0^2\, \left({\bm \lambda} \cdot \meanWW\right) .
\label{B11}
\end{eqnarray}
Therefore, the $\alpha$ effect and the effective  pumping velocity
are caused by a combined effect of the density-stratified turbulence and
a large-scale shear.

We consider the background turbulence being a non-helical, but the resulting turbulence
becomes helical due to the joint effects of the large-scale shear and density stratification.
Indeed, let us determine the kinetic helicity $H_{\rm u}^{(\lambda)} =
\langle {\bm u} {\bf \cdot} (\bec{\nabla} {\bf \times} {\bm u}) \rangle$
in a density-stratified turbulence with a large-scale shear:
$H_{\rm u}^{(\lambda)} = {\rm i} \, \varepsilon_{ijs} \int k_s \, f_{ij}^{(S)}({\bm  k}) \, d{\bm  k} $,
where $f_{ij}^{(S)}({\bm  k})$ is given by equation~(\ref{B4}).
After integration in ${\bm  k}$ space, we obtain that
the kinetic helicity in a density-stratified
turbulence with non-uniform large-scale flow is given by
\begin{eqnarray}
H_{\rm u}^{(\lambda)} = - {5 \over 2} \,  \eta_{_{T}} \,  \, \left({\bm \lambda} \cdot \meanWW\right) ,
\label{B12}
\end{eqnarray}
where $\eta_{_{T}} =(\tau_0 /3)\,  \left\langle {\bm u}^2 \right\rangle$ is the turbulent diffusion coefficient.
Equation for $\eta_{_{T}}$ has been first obtained by the quasi-linear approach (the second-order correlation approximation, SOCA) in high conductivity limit \citep[see, e.g.,][]{Krause(1980)}, and it has been also reproduced using the $\tau$ approaches and the path-integral approach \citep[see, e.g.,][]{RI21}.
Note that the effect of large-scale shear on the turbulent diffusion coefficient
is neglected for a weak large-scale shear ($\meanW \tau_0 \ll 1$).
To derive equations~(\ref{B9})--(\ref{B12}), we used equations~(\ref{B7})--(\ref{MB7}).
Applying the classical expression for the kinetic $\alpha$ effect, we obtain
\begin{eqnarray}
\alpha^{({\rm cl},\lambda)} \equiv - {\tau_0 \over 3} \, H_{\rm u}^{(\lambda)} = {5 \over 18} \, \ell_0^2 \, \left({\bm \lambda} \cdot \meanWW\right) ,
\label{B14}
\end{eqnarray}
which is in a qualitative agreement with equation~(\ref{B11}), but the coefficients in equations~(\ref{B11}) and~(\ref{B14}) do not coincide.
This is not surprising, because the classical expression $\alpha = - (\tau_0 / 3) \, H_{\rm u}$ for the kinetic $\alpha$ effect is only valid for a homogeneous and isotropic helical background turbulence.

The joint action of the $\alpha$ effect and the large-scale shear results in
the generation of the large-scale magnetic field due to
the $\alpha$ - shear mean-field dynamo in the small-scale turbulence
with large fluid and magnetic Reynolds numbers
\citep[see, e.g.,][]{Moffatt(1978),Krause(1980),Zeldovich(1983),RI21}.
Indeed, let us consider the following equilibrium: $\alpha=$ const
and $\overline{\bm U}=(0,Sx,0)$, and take into account
that turbulent diffusion coefficient $\eta_{_{T}} \gg \eta$.
We seek for a solution of the induction equation for perturbations of the large-scale magnetic field as
\begin{eqnarray}
\overline{\bm B}(t,x,z)=\overline{B}_y(t,x,z) {\bm e}_y + {\bm \nabla} {\bm \times}\left[\, \overline{A}(t,x,z) {\bm e}_y\right] ,
\label{N6}
\end{eqnarray}
where ${\bm e}_y$ is the unit vector directed along $y$ axis, and
the functions $\overline{B}_y(t,x,z)$ and $\overline{A}(t,x,z)$ are determined by the following equations:
\begin{eqnarray}
\frac{\partial \overline{A}(t,x,z)}{\partial t} = \alpha \overline{B}_y + \eta_{_{T}} \Delta \overline{A},
\label{N15}
\end{eqnarray}
\begin{eqnarray}
\frac{\partial \overline{B}_y(t,x,z)}{\partial t} = - \alpha \Delta \overline{A} - S \nabla_z \overline{A} + \eta_{_{T}}
\Delta \overline{B}_y ,
\label{N16}
\end{eqnarray}
where $\alpha\equiv \alpha^{(\lambda)}$ is given by equation~(\ref{B11}).
The second term, $- S \nabla_z \overline{A}$, in the right hand side of equation~(\ref{N16})
is originated from the term $(\overline{\bm B} \cdot \bec{\nabla}) \overline{U}_y$.
We consider the case when $|\alpha \Delta \overline{A}| \ll |S \nabla_z \overline{A}|$,
which is valid when $\ell_0^2 \ll L_B \, H_\rho$, where $L_B$ is the characteristic scale
of the mean magnetic field variations and $H_\rho=|{\bm \lambda}|^{-1}$ is the mean density variation scale,
which is assumed to be constant.
The growth rate of the dynamo instability and the frequency of the dynamo waves are given by
\begin{eqnarray}
\gamma = \left({|\alpha  \, S \, K_z| \over 2}\right)^{1/2} - \eta_{_{T}} K^2 ,
\label{N20}
\end{eqnarray}
\begin{eqnarray}
\omega= \left({|\alpha  \, S \, K_z|  \over 2}\right)^{1/2} .
\label{N21}
\end{eqnarray}
The maximum growth rate of the dynamo instability and the maximum frequency
of the dynamo waves are given by
\begin{eqnarray}
\gamma^{\rm max} \approx {3 \over 8} \left({\alpha^2 \, S^2 \over 2\eta_{_{T}}} \right)^{1/3},
\label{N25}
\end{eqnarray}
\begin{eqnarray}
\omega^{\rm max} = {1 \over 2}\left({\alpha^2 \, S^2 \over 2\eta_{_{T}}}
\right)^{1/3} .
\label{N26}
\end{eqnarray}
The maximum growth rate of the dynamo instability and the maximum frequency
of the dynamo waves are attained at $K_x^{\rm max}=0$ and
\begin{eqnarray}
K_z^{\rm max}={1 \over 2} \left({|\alpha \, S| \over 4\eta_{_{T}}^2} \right)^{1/3} .
\label{N24}
\end{eqnarray}
The dynamo instability $\gamma > 0$ implies that
\begin{eqnarray}
{\ell_0 \over L_B }  \left({H_\rho \over L_B}\right)^{1/2} < S \tau_0 \ll 1 .
\label{DN24}
\end{eqnarray}

\section{Inhomogeneous turbulence with large-scale shear}
\label{sect-4}

For comparison, in this section we consider an inhomogeneous and incompressible turbulence with a large-scale shear.
In this case,  the model for the second moment $f_{ij}^{(0)}$  of velocity fluctuations in an inhomogeneous and incompressible
background turbulence in a Fourier space is given by:
\begin{eqnarray}
f_{ij}^{(0)} = {\left\langle {\bm u}^2 \right\rangle^{(0)} \, E(k) \over 8 \pi k^2} \, \biggl[\delta_{ij} - k_{ij}
- {{\rm i} \over 2 k^2} \, \big(\Lambda_i k_j - \Lambda_j k_i\big)\biggr]
\label{C1}
\end{eqnarray}
\citep[see, e.g.,][]{RI21},
where ${\bm \Lambda} = {\bm \nabla} \ln \langle {\bm u}^2 \rangle^{(0)}$.
The contributions to the turbulent electromotive force caused by an inhomogeneous and incompressible turbulence
with a large-scale shear are ${\cal E}_i^{(\Lambda)}
= a_{ij}^{(\Lambda)} \, \meanB_j$, where $a_{ij}^{(\Lambda)}$ is given
by equation~(\ref{MD2}) in Appendix~\ref{Appendix A}.

We also take into account the small-scale dynamo with inhomogeneous magnetic fluctuations in the background turbulence.
To this end, we use the following model for the second moment $b_{ij}^{(0)}({\bm k})$  of magnetic fluctuations:
\begin{eqnarray}
&& b_{ij}^{(0)} = {\left\langle {\bm b}^2 \right\rangle^{(0)} E_{\rm M}(k) \over 8 \pi k^2} \biggl[\delta_{ij} - k_{ij}
- {{\rm i} \over 2 k^2} \big(\Lambda_i^{\rm (M)} k_j
\nonumber\\
&&\quad - \Lambda_j^{\rm (M)} k_i\big)\biggr] ,
\label{IMB7}
\end{eqnarray}
where  ${\bm \Lambda}^{\rm (M)} = {\bm \nabla} \ln \langle {\bm b}^2 \rangle^{(0)}$.
For simplicity, we assume that $E_{\rm M}(k)=E(k)$ for $k \geq \ell_{_{\rm M}}^{-1}$ and $E_{\rm M}(k)=0$
for $\ell_0^{-1}< k<\ell_{\rm M}^{-1}$.

Using this equation, we determine the tensor $\alpha_{ij}^{(\Lambda)} = (a_{ij}^{(\Lambda)} + a_{ji}^{(\Lambda)})/2 $
and the effective  pumping velocity
$V^{\rm eff}_{n}({\bm \Lambda}) = - \varepsilon_{ijn} \, a_{ij}^{(\Lambda)} / 2$ of the mean magnetic field
in an inhomogeneous and incompressible turbulence with a large-scale shear:
\begin{eqnarray}
&& \alpha^{(\Lambda)}_{ij}  = - {\ell_0^2  \over 9} \,
\biggl[\left({\bm \Lambda} \cdot \meanWW\right) \,  \delta_{ij} - {1 \over 2} \Big(\Lambda_i \, \meanW_j + \Lambda_j \, \meanW_i\Big)
\nonumber\\
&& \quad + {1 \over 5}
(4 q - 2) \, \Lambda_m \, \Big(\varepsilon_{imn} \, (\partial \meanU)_{nj}
+ \varepsilon_{jmn} \, (\partial \meanU)_{ni}\Big) \biggr] ,
\label{C3}
\end{eqnarray}

\begin{eqnarray}
&& V^{\rm eff}_{i}({\bm \Lambda}) = - {\ell_0^2 \over 18} \, \biggl[
 \left({\bm \Lambda} \times \meanWW \right)_{i}
- 4 \Lambda_m (\partial \meanU)_{mi}\biggr] .
\label{C4}
\end{eqnarray}

Equations~(\ref{C3})--(\ref{C4}) are given in the absence of the small-scale dynamo.
The tensor $\alpha_{ij}^{(\Lambda_{\rm M})}  = (a_{ij}^{(\Lambda_{\rm M})} + a_{ji}^{(\Lambda_{\rm M})})/2 $ and the effective  pumping velocity $V^{\rm eff}_{n}({\bm \Lambda}_{\rm M}) = - \varepsilon_{ijn} \, a_{ij}^{(\Lambda_{\rm M})} / 2$
caused by the small-scale dynamo are given by
\begin{eqnarray}
&& \alpha^{(\Lambda_{\rm M})}_{ij}  = - {4(q-1)  \over 45} \,\ell_0^2 \, \, \biggl({\ell_{_{\rm M}} \over \ell_0}\biggr)^{3(q-1)}
\biggl[{ \left\langle {\bm b}^2 \right\rangle^{(0)} \over \mu_0 \meanrho \, \left\langle {\bm u}^2 \right\rangle^{(0)} } \biggr]
\nonumber\\
&& \quad
\times \Lambda_m^{\rm (M)} \, \Big(\varepsilon_{imn} \, (\partial \meanU)_{nj}
+ \varepsilon_{jmn} \, (\partial \meanU)_{ni}\Big)  ,
\label{MC3}
\end{eqnarray}
\begin{eqnarray}
&& V^{\rm eff}_{i}({\bm \Lambda}_{\rm M}) = {\ell_0^2 \over 45}  \, \, \biggl({\ell_{_{\rm M}} \over \ell_0}\biggr)^{3(q-1)}
\biggl[{ \left\langle {\bm b}^2 \right\rangle^{(0)} \over \mu_0 \meanrho \, \left\langle {\bm u}^2 \right\rangle^{(0)} } \biggr]
\nonumber\\
&& \quad
\times \biggl[5 \left({\bm \Lambda}^{\rm (M)} \times \meanWW \right)_{i}
+ 2 (2q-7)\Lambda_m^{\rm (M)} (\partial \meanU)_{mi}\biggr] ,
\label{MC4}
\end{eqnarray}
where $a_{ij}^{(\Lambda_{\rm M})}$ is given by equation~(\ref{MBAD4}) in Appendix~\ref{Appendix A}.

The kinetic $\alpha$  effect based on the isotropic part ($\propto \delta_{ij}$) of the alpha tensor
in an inhomogeneous turbulence with a large-scale shear is given by
\begin{eqnarray}
\alpha^{(\Lambda)} = - {\ell_0^2 \over 9} \, \left({\bm \Lambda} \cdot \meanWW\right) .
\label{C5}
\end{eqnarray}
Therefore, the $\alpha$ effect and the effective  pumping velocity
are caused by a joint effect of the inhomogeneous turbulence and
a nonuniform large-scale flow.
Equations~(\ref{C3})--(\ref{C4})  are in agreement with those derived by
\cite{RK03} using the spectral $\tau$ approach and by \cite{RS06}
applied the quasi-linear approach for high conductivity limit but small Strouhal numbers.

Note that equations~(\ref{C3})--(\ref{C4}) for
the tensor $\alpha_{ij}^{(\Lambda)}$ and the effective  pumping velocity
${\bm V}^{\rm eff}({\bm \Lambda})$
are different from equations~(\ref{B9})--(\ref{B10})
derived for a density-stratified turbulence with a large-scale shear.
The reason is caused by a difference between the effects of the density stratification and the inhomogeneity of turbulence
on the tensor $\alpha_{ij}$ and the effective  pumping velocity ${\bm V}^{\rm eff}$.
Indeed, the density stratification affects
\begin{itemize}
\item{
the background turbulence [see see equations~(\ref{B7}) and~(\ref{QB6})];}
\item{
the tensors $I_{ijmn}(\meanUU)$ and $J_{ijmn}(\meanUU)$,
which describe the effect of the large-scale shear on turbulence [see equations~(\ref{B5}) and~(\ref{B6})];}
\item{
and the term $- \meanBB \, {\rm div} {\bm  u}$ in the induction equation ~(\ref{B1})
which causes the appearance of the term $- \lambda_n \, \overline{B}_j \, f_{in}^{(S)}$ in equation ~(\ref{B3})
for the cross-helicity tensor $g_{ij}^{(S)}({\bm  k})$.}
\end{itemize}
On the other hand,  the parameter $\Lambda_i$ characterising the inhomogeneous turbulence
affects only the background turbulence [see see equations~(\ref{C1}) and~(\ref{QB6})].
This causes the difference between the effects of the density stratification and the inhomogeneity of turbulence
on the tensor $\alpha_{ij}$ and the effective  pumping velocity $V^{\rm eff}_{i}$ [compare equations~(\ref{B9})--(\ref{B10})
with~(\ref{C3})--(\ref{C4})].

We consider the background non-helical turbulence, but the resulting turbulence
becomes helical due to the joint effects of the large-scale shear and inhomogeneity
of turbulence.
Indeed, let us determine the kinetic helicity $H_{\rm u}^{(\Lambda)} =
\langle {\bm u} {\bf \cdot} (\bec{\nabla} {\bf \times} {\bm u}) \rangle$
in an inhomogeneous turbulence with a large-scale shear.
Using equations~(\ref{B4}) and~(\ref{C1}), and integrating in ${\bm  k}$ space in
expression $H_{\rm u}^{(\Lambda)} = {\rm i} \, \varepsilon_{ijs} \int k_s \, f_{ij}^{(S)}({\bm  k}) \, d{\bm  k} $,
we obtain that the kinetic helicity in an inhomogeneous
turbulence with a large-scale shear is given by
\begin{eqnarray}
H_{\rm u}^{(\Lambda)} = \eta_{_{T}} \, \left({\bm \Lambda} \cdot \meanWW\right) ,
\label{C6}
\end{eqnarray}
where $\eta_{_{T}}$ is the turbulent diffusion coefficient.
Applying a classical expression for the kinetic $\alpha$ effect, $\alpha^{({\rm cl},\Lambda)}
\equiv - (\tau_0 /3)\, H_{\rm u}^{(\Lambda)}$, we obtain that
\begin{eqnarray}
\alpha^{({\rm cl},\Lambda)} = - {\ell_0^2 \over 9} \, \left({\bm \Lambda} \cdot \meanWW\right) ,
\label{C7}
\end{eqnarray}
which coincides with equation~(\ref{C5}).
May be it is due to the fact that inhomogeneity of turbulence
is a more simple effect that only is determined by the background turbulence.

\section{Compressible turbulence with large-scale shear}
\label{sect-5}

In this section we consider five simple independent effects:
\begin{itemize}
\item{
the stratification of a small-scale background turbulence described by the parameter ${\bm \lambda}$
(div ${\bm u} = {\bm \lambda} \cdot {\bm u})$;}
\item{
the finite Mach number effects for a compressible small-scale background turbulence
described by the parameter $\sigma_c$;}
\item{
the imposed large-scale shear described by a non-zero mean vorticity $\meanWW =$ rot $\meanUU$;}
\item{
the imposed mean fluid motion with a non-zero div $\meanUU$, which
allows to describe collapsing (or expanding) astrophysical clouds;}
\item{
the inhomogeneity of a small-scale background turbulence
described by the parameter ${\bm \Lambda}$.}
\end{itemize}
For simplicity, we assume that these five given parameters are independent.
We investigate how these independent parameters affect
the kinetic ${\bm \alpha}$ tensor and the effective  pumping velocity.
Generally, the mean velocity field $\meanUU$ is determined by
the mean Navier-Stokes equation and the mean continuity equation.

The tensor $f_{ij}^{(0)}({\bm k})$ for a density-stratified, inhomogeneous and
compressible non-helical background turbulence for arbitrary Mach numbers in the ${\bm k}$ space
is given by
\begin{eqnarray}
f_{ij}^{(0)} &=&
{\left\langle {\bm u}^2 \right\rangle^{(0)} \,  E(k) \over 8 \pi \, k^2} \,
\biggl[ (\delta_{ij} - k_{ij} + 2 \sigma_c \, k_{ij}) \, (1+ \sigma_c)^{-1}
\nonumber\\
&& + {{\rm i} \over k^2}  \, \biggl(\lambda_i k_j  - \lambda_j k_i
+ {1 \over 2} \, \big(k_i \Lambda_j - k_j \Lambda_i\big) \biggr) \biggr]
\label{QB6}
\end{eqnarray}
\citep[see, e.g.,][]{RI21},
where the parameter
\begin{eqnarray}
\sigma_c = {\left\langle (\bec{\nabla} \cdot \, {\bm u})^2 \right\rangle
\over \left\langle(\bec{\nabla} \times {\bm u})^{2} \right\rangle}
  \label{RB40}
\end{eqnarray}
describes the degree of compressibility of a turbulent velocity field.
The background turbulence model given by equation~(\ref{QB6})
is derived from the symmetry arguments under
the condition $\ell_0 \ll H_\rho$ and $\ell_0 \ll L_u$.
Here $L_u= |{\bm \Lambda}|^{-1}=\left|\bec{\nabla}
\ln \left\langle {\bm u}^2\right\rangle^{(0)}\right|^{-1}$
is the characteristic scale of the inhomogeneity of turbulence, and
$H_\rho=|{\bm \lambda}|^{-1}$ is the mean density variation scale,
which is assumed to be constant.
This implies that we use the perturbation approach, i.e.,
equation~(\ref{QB6}) takes into account leading-order effects, which are linear
in stratification ($\propto \ell_0 / H_\rho$) and inhomogeneity of turbulence
($\propto \ell_0 / L_u$), and the higher-order effects
$\sim$O$(\ell_0^2 / H_\rho^2, \ell_0^2 /L_u^2)$ are neglected.

Generally, stratification also contributes to the parameter $\sigma_c$.
However, this contribution is small [$\sim$O$(\ell_0^2 / H_\rho^2)$],
and neglected in equation~(\ref{QB6}).
This implies that the effects of the arbitrary Mach number, characterized by
the parameter $\sigma_c$, and density stratification, described by ${\bm \lambda}$
are separated.
The degree of compressibility $\sigma_c$ depends on the Mach number, but an
analytical dependence $\sigma_c$ on the Mach number is not known for arbitrary Mach numbers
and it can be determined in numerical simulations.
For small Mach numbers ${\rm Ma} \ll 1$, the parameter $\sigma_c \sim {\rm Ma}^5 {\rm Re}^{1/4}$
\citep{RK21B,RK21A},
where ${\rm Re}$ is the Reynolds number based on the integral scale and turbulent velocity.

Since we consider only linear effects in ${\bm \lambda}$ and ${\bm \Lambda}$,
the tensor $f_{ij}^{(0)}$ is constructed as a linear combination of symmetric tensors,
$\delta_{ij}$ and $k_{ij}$, with respect to the indexes $i$ and $j$, and non-symmetric tensors,
$k_i \lambda_j$, $ k_j \lambda_i$, and $k_i \Lambda_j$, $ k_j \Lambda_i$.
To determine unknown coefficients multiplying by these tensors,
we use the following conditions in the derivation of Eq.~(\ref{QB6}):
\begin{eqnarray}
\left\langle {\bm u}^2 \right\rangle^{(0)} = \int f_{ii}^{(0)} ({\bm k},{\bm R}) \, d {\bm k} ,
 \label{RZ1}
\end{eqnarray}
\begin{eqnarray}
\left\langle\left({\rm div} \, {\bm u}\right)^2\right\rangle  = \int k_i \, k_j \, f_{ij}^{(0)} ({\bm k},{\bm R}) \, d {\bm k}  ,
 \label{RZ4}
\end{eqnarray}
\begin{eqnarray}
\left\langle\left({\rm rot} \, {\bm u}\right)^2\right\rangle  =\int k^2 \, f_{ii}^{(0)} ({\bm k},{\bm R})\, d {\bm k}
- \left\langle\left({\rm div} \, {\bm u}\right)^2\right\rangle ,
 \label{NRZ4}
\end{eqnarray}
where ${\bm R}$ corresponds to large scales.

We assume here that the background turbulence is of Kolmogorov type with
constant energy flux over the spectrum,
i.e., the turbulent kinetic energy spectrum in the range of wave numbers $k_0<k<k_\nu$ is
$E(k) = - d \bar \tau(k) / dk$, where the function $\bar \tau(k) =
(k / k_{0})^{1-q}$ with $1 < q < 3$
\citep[see, e.g.,][]{RI21}.
The exponent $q=5/3$ corresponds to the Kolmogorov spectrum,
while the exponent $q=2$ corresponds to the spectrum of the Burgers turbulence.
The turbulent time in the ${\bm k}$ space is
$\tau(k) = 2\tau_0 \, \bar \tau(k)$.

Let us first consider a density-stratified and inhomogeneous non-helical
background turbulence with very small degree of compressibility $\sigma_c \ll 1$.
In this case, the total kinetic helicity $H_u^{\rm (tot)} = H_u^{(\lambda)} + H_u^{(\Lambda)}$
is given by
\begin{eqnarray}
H_u^{\rm (tot)} = 2 \eta_{_{T}} \, \left(\meanWW \cdot {\bm \nabla} \right)
\ln \left(\meanrho^{\,5/4} \, u_{\rm rms} \right) ,
\label{HC8}
\end{eqnarray}
where $u_{\rm rms}$ is defined as $u_{\rm rms} = \sqrt{\langle{\bm u}^2\rangle}$,
and the angular brackets $\langle ... \rangle$ denote the ensemble averaging.
The total kinetic
$\alpha$ tensor is given by $\alpha_{ij} = \alpha_{ij}^{(\lambda)} + \alpha_{ij}^{(\Lambda)}$
and the effective  pumping velocity is
$V^{\rm eff}_{i} = V^{\rm eff}_{i}({\bm \lambda}) + V^{\rm eff}_{i}({\bm \Lambda})$.
In this case,  the kinetic $\alpha$  effect based on the isotropic part ($\propto \delta_{ij}$) of the kinetic $\alpha$ tensor is given by
\begin{eqnarray}
\alpha^{\rm (tot)} = - {2 \over 9} \, \ell_0^2 \,\left(\meanWW \cdot {\bm \nabla} \right)
\ln \left(\meanrho^{\,13/10} \, u_{\rm rms} \right) ,
\label{C8}
\end{eqnarray}
where we used equations~(\ref{B11}) and~(\ref{C5}).
On the other hand, the classical expression for the kinetic $\alpha$ effect, $\alpha^{({\rm cl},{\rm tot})}
= - (\tau_0 /3)\, H_{\rm u}^{\rm tot}$,
is given by
\begin{eqnarray}
\alpha^{({\rm cl},{\rm tot})} = - {2 \over 9} \, \ell_0^2 \, \, \left(\meanWW \cdot {\bm \nabla} \right)
\ln \left(\meanrho^{\, 5/4} \, u_{\rm rms} \right) ,
\label{C9}
\end{eqnarray}
where we used equations~(\ref{B14}) and~(\ref{C7}).
Equations~(\ref{C8}) and~(\ref{C9}) are not coincided.
This is not surprising, because the classical expression $\alpha = - (\tau_0 / 3) \, H_{\rm u}$
for the kinetic $\alpha$ effect is only valid for a homogeneous and isotropic helical turbulence.

Let us compare equation~(\ref{C8}) for turbulence with a large-scale shear with that for a slowly rotating turbulence ($\meanOmega \tau_0 \ll 1$), where the following scaling for the $\alpha$ effect have been obtained in different studies:
\begin{eqnarray}
\alpha^{(\Omega)} \propto - \ell_0^2 \, \, \left(\meanOmega \cdot {\bm \nabla} \right)
\ln \left(\meanrho^{\, \mu_\ast} \, u_{\rm rms} \right) ,
\label{RC9}
\end{eqnarray}
with $\mu_\ast =1$ \citep{Steenbeck(1966),Krause(1980)} by means of the quasi-linear approach, $\mu_\ast =3/2$
\citep{RUK93} applying the modified quasi-linear approach, and $\mu_\ast =1/2$ \citep{BGKR13} using the spectral $\tau$
approach. Here $\meanOmega$ is the mean angular velocity describing a uniform rotation.

Note that the cases of uniform rotation and large-scale shear
with a density-stratified or inhomogeneous turbulence
are physically two different cases.
However, this comparison shows that in both cases
(uniform rotation and large-scale shear) with
density-stratified or inhomogeneous turbulence,
there is a production of  the kinetic helicity and the $\alpha$ effect,
and the form of the alpha effect are similar in both cases
with the replacement the angular velocity $\meanOmega$
by the large-scale vorticity $\meanWW$.

The physics of this effect is the following.
Both, the angular velocity $\meanOmega$ and the large-scale vorticity $\meanWW$ produce
the left-handed and right-handed rotating turbulent eddies.
A non-zero kinetic helicity implies that a  number of the left-handed eddies
at a given instant does not exactly equal the number of right-handed eddies.
This breaking a symmetry between the numbers of the left-handed and right-handed turbulent eddies
is caused by density-stratified or inhomogeneous turbulence.
A mechanism of the $\alpha$ effect is as follows. Deformation of the original magnetic field
is caused by both, the left-handed and the right-handed rotating eddies.
Due to the breaking a symmetry, the total effect of
deformation of the original magnetic field line is not zero,
which causes the generation of a large-scale magnetic field.

Now we consider a density-stratified,  homogeneous and compressible
turbulence.
In this case,
the kinetic ${\bm \alpha}$ tensor $\alpha_{ij}^{(\lambda,\sigma_c)}  = (a_{ij}^{\rm (tot)} + a_{ji}^{\rm (tot)})/2 $ and the effective  pumping velocity $V^{\rm eff}_{n}({\bm \lambda},\sigma_c) = - \varepsilon_{ijn} \, a_{ij}^{\rm (tot)} / 2$ of the mean magnetic field are given by
\begin{eqnarray}
&& \alpha_{ij}^{(\lambda,\sigma_c)}  = - {\ell_0^2 \over 45}  \,
\biggl\{\biggl(2 +{15 \sigma_c \over 2(1 + \sigma_c)} \biggr) \,
\Big(\lambda_i \, \meanW_j+ \lambda_j \, \meanW_i\Big)
\nonumber\\
&& \quad - 13 ({\bm \lambda} \cdot \meanWW)  \,\delta_{ij} + \biggl[\biggl(
4q-7 -{6\sigma_c \over 1 + \sigma_c} \biggr)
\Big(\varepsilon_{imn} \, (\partial \meanU)_{nj}
\nonumber\\
&& \quad
+ \varepsilon_{jmn} \, (\partial \meanU)_{ni}\Big) \lambda_m \biggr\} ,
\label{WB1}
\end{eqnarray}

\begin{eqnarray}
&& V^{\rm eff}_{i}(\lambda,\sigma_c) = {\ell_0^2 \over 45} \, \biggl[5 \biggl(1 -
{\sigma_c \over 2 (1 + \sigma_c)} \biggr) \Big({\bm \lambda}
\times \meanWW \Big)_i
\nonumber\\
&& \quad + 2 \biggl(
8q +11 -(12q + 13)
{\sigma_c \over 1 + \sigma_c} \biggr) \lambda_m (\partial \meanU)_{mi}
\nonumber\\
&& \quad
+4 \biggl(
11  -7 q +(2q -3) \, {\sigma_c \over 1 + \sigma_c} \biggr) \lambda_i  {\rm div} \meanUU \biggr] ,
\label{WB4}
\end{eqnarray}
where $a_{ij}^{\rm (tot)}$ is given
by equation~(\ref{MD3}) in Appendix~\ref{Appendix A}.
Equations~(\ref{WB1})--(\ref{WB4}) are obtained for homogeneous turbulence
$({\bm \Lambda}=0)$.
For an inhomogeneous, nonstratified and compressible turbulence (with a nonzero parameter $\sigma_c$, i.e., in turbulence
with a finite Mach numbers),
the kinetic ${\bm \alpha}$ tensor is independent of the Mach number
[i.e., is independent of the parameter $\sigma_c$ and is given by equation~(\ref{C3})].
On the other hand,
the effective  pumping velocity $V^{\rm eff}_{n}({\bm \Lambda},\sigma_c)$ of the mean magnetic field
depends on the Mach number (i.e., it depends on the parameter $\sigma_c$):
\begin{eqnarray}
&& V^{\rm eff}_{i}({\bm \Lambda},\sigma_c) = {2 \ell_0^2 \over 45} \, \biggl[
\biggl(
5 - 2(2q + 1) {\sigma_c \over 1 + \sigma_c}\biggr) \Lambda_m  (\partial \meanU)_{mi}
\nonumber\\
&&
- {5 \over 4} \Big({\bm \Lambda}  \times \meanWW\Big)_i
- \biggl(4q+5
+ (2q + 1)  {\sigma_c \over 1 + \sigma_c}\biggr) \Lambda_i  \, {\rm div} \meanUU
\biggr] .
\nonumber\\
\label{CWB4}
\end{eqnarray}
Equation~(\ref{CWB4}) is obtained for non-stratified turbulence
$({\bm \lambda}=0)$.

Note that the parameter $\sigma_c$ does not affect the terms $\propto \lambda_i$
and $\propto  \Lambda_i$ in equation~(\ref{QB6}), that takes into account only the leading-order effects.
However, the parameter $\sigma_c$ affects the contributions caused by the density stratifications
to the the ${\bm \alpha}$ tensor and effective pumping velocity ${\bm V}^{\rm eff}$,
because the density stratifications influence the large-scale shear contributions
[see comment after equation~(\ref{C5})].

In this study we also take into account a possibility for collapsing (or expanding) astrophysical clouds,
which can be described by a non-zero div $\meanUU$.
This implies that we consider a large-scale dynamo with a large-scale shear (a non-zero
large-scale vorticity $\meanWW$) and  collapsing (or expanding) large-scale motions with
a non-zero div $\meanUU$.
This effect causes new contributions to the effective pumping velocity of the mean magnetic field
${\bm V}^{\rm eff} \propto {\bm \lambda} \,  {\rm div}\, \meanUU$
in density stratified turbulence, or ${\bm V}^{\rm eff} \propto {\bm \Lambda} \,  {\rm div} \, \meanUU$
in inhomogeneous turbulence, which can arise in collapsing (or expanding) astrophysical turbulent clouds.
However, the $\alpha_{ij}$ tensor is independent of ${\rm div} \, \meanUU$, i.e.,
it is independent of the effects of collapsing or expanding of clouds.
The isotropic part of the ${\bm \alpha}$ tensor ($\propto \delta_{ij}$) is independent of the exponent
$q$ of the turbulence energy spectrum.

For illustration various contributions
to the ${\bm \alpha}$ tensor and effective pumping velocity ${\bm V}^{\rm eff}$,
we consider a small-scale turbulence with large-scale linear
velocity $\meanUU=(a_{_{U}} x/3, Sx + a_{_{U}} y/3, a_{_{U}} z/3)$ in the Cartesian coordinates $(x,y,z)$,
where the large-scale vorticity is $\meanWW=(0, 0, S)$
and ${\rm div} \, \meanUU = a_{_{U}}$.
The stress tensor $(\partial \meanU)_{ij} = (S/2) \, (e_i^x \, e_j^y + e_j^x \, e_i^y) + a_{_{U}} \delta_{ij}/3$,
where ${\bm e}^x$, ${\bm e}^y$ and ${\bm e}^z$ are the unit vectors.
The vector ${\bm \lambda}$ that describes the stratification of the mean fluid density, is
${\bm \lambda}=\lambda \, (0, 0, 1)$,
and the vector ${\bm \Lambda}$ that determines the inhomogeneity of turbulence, is
${\bm \Lambda}=\Lambda \, (0, 0, 1)$.
The tensor $C_{ij}^{(\lambda)} =\lambda_m  \, [\varepsilon_{imn} \, (\partial \meanU)_{nj}+ \varepsilon_{jmn} \, (\partial \meanU)_{ni}]$ entering in equation~(\ref{WB1}),
has the following diagonal components: $C_{xx}^{(\lambda)} = - 2\lambda (\partial \meanU)_{xy} = - \lambda \, S$, $C_{yy}^{(\lambda)} = 2\lambda (\partial \meanU)_{xy} = \lambda \, S$ and $C_{zz}^{(\lambda)}=0$.
This yields the following diagonal components of the ${\bm \alpha}$ tensor:
\begin{eqnarray}
\alpha^{(\lambda,\Lambda)}_{xx} = -{2(4 q + 7) \over 45}  \, \ell_0^2 \, S \, \nabla_z
\ln \Big[\meanrho^{\, \mu_1} u_{\rm rms} \Big]   ,
\label{WA10}
\end{eqnarray}

\begin{eqnarray}
\alpha^{(\lambda,\Lambda)}_{yy} = - {2(4 q +3)  \over 45}  \, \ell_0^2 \,S \, \nabla_z
\ln \Big[\meanrho^{\, \mu_2} u_{\rm rms} \Big]   ,
\label{WB10}
\end{eqnarray}

\begin{eqnarray}
\alpha^{(\lambda,\Lambda)}_{zz} = -{2 \over 9}  \, \ell_0^2 \, S \, \nabla_z
\ln \Big[\meanrho^{\, \mu_3} u_{\rm rms} \Big]   ,
\label{WC10}
\end{eqnarray}
where
\begin{eqnarray}
\mu_1 = - {1 \over 4 q + 7 } \, \left(2 q +3 -  {3 \sigma_c \over 1 + \sigma_c} \right)  ,
\label{WA11}
\end{eqnarray}

\begin{eqnarray}
\mu_2 = {1 \over 4 q +3 } \, \left(10 - 2q + {3 \sigma_c \over 1 + \sigma_c} \right)  .
\label{WB11}
\end{eqnarray}
and $\mu_3 = 13/10$.
The effective pumping velocity ${\bm V}^{\rm eff}$ is given by
\begin{eqnarray}
V^{\rm eff}_{x} = - {2 \over 9} \biggl[1- {2 \over 5} (2q +1) {\sigma_c \over 1 + \sigma_c}\biggr]  \ell_0^2 \, S \, \nabla_z \ln \Big[\meanrho^{\, \mu_x} u_{\rm rms} \Big]  ,
\nonumber\\
\label{XWC10}
\end{eqnarray}
\begin{eqnarray}
V^{\rm eff}_{y} ={2 \over 9} \biggl[1- {2 \over 5} (2q +1) {\sigma_c \over 1 + \sigma_c}\biggr] \ell_0^2 \, S \,  \nabla_z \ln \Big[\meanrho^{\, \mu_y} \, u_{\rm rms} \Big]  ,
\nonumber\\
\label{YWC10}
\end{eqnarray}
where
\begin{eqnarray}
\mu_y &=&  - 2 \mu_x = - \biggl[8q + 11 - (12 q + 13) \, {\sigma_c \over 1 + \sigma_c} \biggr]\,
\nonumber\\
&& \times
\biggl[5 -2 (2q + 1) {\sigma_c \over 1 + \sigma_c} \biggr] ^{-1} .
\label{XWB15}
\end{eqnarray}
The contribution to the effective pumping velocity of the mean magnetic field
caused by collapsing (or expanding) clouds described by the divergence of the mean fluid velocity
is given by
\begin{eqnarray}
V^{\rm eff}_{z} &=& - {8 \over 21} \ell_0^2 \, {\rm div} \meanUU
\biggl[{6q \over 5}+1+ \Big(q + {1 \over 2}\Big)  {\sigma_c \over 1 + \sigma_c}\biggr]
\nonumber\\
&& \times
\nabla_z \ln \Big[\meanrho^{\, \mu_z} \, u_{\rm rms} \Big]  ,
\label{ZWB30}
\end{eqnarray}
where
\begin{eqnarray}
\mu_z &=& {1 \over 20}\biggl[77 -34 q - 31 \, {\sigma_c \over 1 + \sigma_c} \biggr]
\nonumber\\
&& \times
\biggl[{6q \over 5}+1+ \Big(q + {1 \over 2}\Big)  {\sigma_c \over 1 + \sigma_c} \biggr] ^{-1} .
\label{YWB15}
\end{eqnarray}

\section{Applications to protoplanetary discs and astrophysical clouds}
\label{sect-6}

In this section, we consider applications of the obtained results
related to the ${\bm \alpha}$ tensor and effective pumping velocity ${\bm V}^{\rm eff}$
to protoplanetary discs and astrophysical clouds.
For simplicity, we consider  here the background turbulence without small-scale dynamo.

\subsection{Protoplanetary disks}
\label{subsect-6.1}

In this section we determine  the ${\bm \alpha}$ tensor and effective pumping velocity ${\bm V}^{\rm eff}$
in protoplanetory disks.
We use the cylindrical coordinates $(r, \varphi, z)$
with corresponding units vectors ${\bm e}^r$, ${\bm e}^\varphi$ and ${\bm e}^z$
along these axes.
We consider a small-scale turbulence with
the large-scale nonuniform axisymmetric velocity $\meanUU=(0, r \, \delta\Omega(r), 0)$,
where $\delta\Omega$ describes differential rotation.
In this case, the large-scale vorticity is $\meanWW = {\bm e}^z \, r^{-1} \, (\partial/\partial r) \, (r^2 \delta\Omega) $.
Thus, $(\partial \meanU)_{r\varphi} = (r/2) \, (\partial/\partial r) \, \delta\Omega$.
The vector ${\bm \lambda}$ that describes the non-uniform mean fluid density, is
${\bm \lambda}=(\lambda_r, 0, \lambda_z)$,
and the vector ${\bm \Lambda}$ that determines the inhomogeneity of turbulence, is
${\bm \Lambda}=(\Lambda_r, 0, \Lambda_z)$.
Thus, we obtain that the $\alpha_{\varphi\varphi}$ component of the ${\bm \alpha}$ tensor
is given by
\begin{eqnarray}
\alpha_{\varphi\varphi} = - {4 \over 9} \, \Big[1 + {{\rm D}_r  \over 10} (4q+3)\Big]   \, \ell_0^2 \, \delta\Omega(r)
\nabla_z \ln \left(\meanrho^{\mu_\alpha} \, u_{\rm rms} \right) ,
\label{DWB14}
\end{eqnarray}
where
\begin{eqnarray}
\mu_\alpha &=& {13 \over 10} \,\biggl[1 + {{\rm D}_r  \over 13}  \Big(10-2q + {3\sigma_c \over 1 + \sigma_c} \Big) \biggr]\,
\nonumber\\
&& \times \Big(1 + {{\rm D}_r  \over 10} (4q+3) \Big)^{-1} ,
\label{DWB15}
\end{eqnarray}
and the parameter characterising the differential rotation defined as
\begin{eqnarray}
{\rm D}_r = {\partial \ln \delta\Omega \over \partial  \ln r} .
\label{DWB16}
\end{eqnarray}
The $\varphi$ component of the effective pumping velocity is
\begin{eqnarray}
V^{\rm eff}_{\varphi} &\approx& {2\ell_0^2  \over 9} \delta\Omega(r)
\, \biggl[1 + {{\rm D}_r \over 10}\biggl(15 - 4 (2q + 1) {\sigma_c \over 1 + \sigma_c} \biggr) \biggr]
\nonumber\\
&& \times \nabla_r \ln \left(\meanrho^{\mu_{\rm v}} \, u_{\rm rms} \right) ,
\label{DWB17}
\end{eqnarray}
where
\begin{eqnarray}
\mu_{\rm v} &=& \biggl[1 - {\sigma_c \over 10(1 + \sigma_c)} - {{\rm D}_r \over 5}  \biggl(3+ 2q - \Big(6 q - {25 \over 4}\Big)
\nonumber\\
&& \times {\sigma_c \over 1 + \sigma_c}   \biggr)  \biggr]\,
\biggl[1 + {{\rm D}_r \over 10} \biggl(15 -4 (2q + 1) {\sigma_c \over 1 + \sigma_c}\biggr) \biggr] ^{-1} ,
\nonumber\\
\label{DWB18}
\end{eqnarray}
To derive equations~(\ref{DWB14}) and~(\ref{DWB17}), we use equations~(\ref{WB1}) and~(\ref{WB4}).

The analyzed effects are important for generation of large-scale magnetic
fields in protoplanetary discs (PPD).
The typical parameters of the protosolar nebula
\citep[see, e.g.,][]{HB98,EKR98,PPK11,HUB16,HOP16,HOP16b,KR25} are as follows:
the angular velocity $\Omega \sim 2 \times 10^{-7} \, r_{_{\rm AU}}^{-3/2} \, {\rm s}^{-1}$
(where $r_{_{\rm AU}}$ is the radial coordinate
measured in the astronomical units $L_{_{\rm AU}}=1.5 \times 10^{13}$ cm);
the shear $\delta\Omega \sim 5 \times 10^{-8} \, r_{_{\rm AU}}^{-5/2} \, {\rm s}^{-1}$;
the sound speed $c_{\rm s} = 6.4 \times 10^{4} \, r_{_{\rm AU}}^{-3/14}$  cm/s;
the integral scale of turbulence $\ell_0 = \sqrt{\alpha_{_{\rm PPD}}} \, c_{\rm s}/\Omega =
3 \times 10^{10} \, r_{_{\rm AU}}^{9/7}$ cm;
the turbulent velocity $u_0 = \alpha_{_{\rm PPD}} \, c_{\rm s} \approx (65$ -- $650) \, r_{_{\rm AU}}^{-3/14}$ cm/s;
the turbulent time $\tau_0=\ell_0/u_0 = (\sqrt{\alpha_{_{\rm PPD}}} \, \Omega)^{-1}$,
the kinematic viscosity $\nu=c_{\rm s} \, \lambda_{\rm mfp} /2 = 1.6 \times 10^{5} \, r_{_{\rm AU}}^{18/7}$ cm$^2$/s,
so the Reynolds number ${\rm Re} = \ell_0 \, u_0 / \nu$ varies in the range
${\rm Re} =(10^{6}$-- $10^{8}) \, r_{_{\rm AU}}^{-3/2}$.
Here $\lambda_{\rm mfp} = 5 \, r_{_{\rm AU}}^{39/14}$ cm is the mean-free path of the gas molecules,
and parameter $\alpha_{_{\rm PPD}}$ varies from $10^{-3}$ to $10^{-2}$.
The mean fluid density is $\meanrho = 2 \times 10^{-9} \, r_{_{\rm AU}}^{-11/4}$ g/cm$^3$
and the density stratification scale $H_{\rm g}=c_{\rm s}/\Omega=3 \times 10^{11} \, r_{_{\rm AU}}^{9/7}$ cm.
This implies that the $\alpha_{\varphi\varphi}$ component of the ${\bm \alpha}$ tensor
is estimated as $|\alpha_{\varphi\varphi}| \approx 10 \, r_{_{\rm AU}}^{-17/14}$ cm/s
and the $\varphi$ component of the effective pumping velocity is
$|V^{\rm eff}_\varphi | \approx 15 \, r_{_{\rm AU}}^{-17/14}$ cm/s.

\subsection{Colliding protogalactic clouds and merging protostellar clouds}
\label{subsect-6.2}

Next, we consider astrophysical clouds, and
use the spherical coordinates $(r, \theta, \varphi)$
with corresponding units vectors ${\bm e}^r$, ${\bm e}^\theta$ and ${\bm e}^\varphi$
along these axes.
This may have relevance to colliding protogalactic clouds (PGC) and merging protostellar clouds (PSC).
Interaction of the merging clouds causes large-scale shear motions which are superimposed on small-scale turbulence.
We consider a small-scale turbulence with
the large-scale nonuniform axisymmetric velocity $\meanUU=(0, 0, r \, \sin \theta\, \delta\Omega(r,\theta))$,
where $\delta\Omega$ determines differential rotation.
Thus, the large-scale vorticity is $\meanWW = {\bm e}^r \, (\sin \theta)^{-1} \, (\partial/\partial \theta) \, (\sin^2 \theta \, \delta\Omega)
+ {\bm e}^\theta \, \sin \theta \, r^{-1} \, (\partial/\partial r) \, (r^2 \delta\Omega) $.
Therefore, $(\partial \meanU)_{\theta\varphi} =(\sin \theta/2) \, (\partial/\partial \theta) \, \delta\Omega$.
The vector ${\bm \lambda}$ that describes the non-uniform mean fluid density, is
${\bm \lambda}=\lambda (1, 0, 0)$,
and the vector ${\bm \Lambda}$ that determines the inhomogeneity of turbulence, is
${\bm \Lambda}=\Lambda (1, 0, 0)$.
Thus, the $\alpha_{\varphi\varphi}$ component of the ${\bm \alpha}$ tensor
is given by
\begin{eqnarray}
\alpha_{\varphi\varphi} &=& - {4 \over 9} \Big[1 + {{\rm D}_\theta \over 10}
(7 - 4q)\Big]   \, \ell_0^2 \, \delta\Omega(r) \, \cos \theta
\nonumber\\
&& \times  \nabla_r \ln \left(\meanrho^{\, \mu_\alpha} \, u_{\rm rms} \right) ,
\label{CWB20}
\end{eqnarray}
where
\begin{eqnarray}
\mu_\alpha &=& {13 \over 10} \, \biggl[1 + {{\rm D}_\theta \over 26} \biggl(
5 +4q -{6\sigma_c \over 1 + \sigma_c} \biggr) \biggr]
\nonumber\\
&& \times  \biggl[1 + {{\rm D}_\theta \over 10}
(7 - 4q)\biggr]^{-1} ,
\label{WB21}
\end{eqnarray}
and the parameter ${\rm D}_\theta$ characterising the latitudinal differential rotation, is defined as
\begin{eqnarray}
{\rm D}_\theta = \tan \theta \, {\partial \over \partial  \theta} \ln \delta \Omega  .
\label{WB22}
\end{eqnarray}
The $\varphi$ component of the effective pumping velocity is
\begin{eqnarray}
V^{\rm eff}_{\varphi} &\approx& {2\ell_0^2  \over 9} \, \delta\Omega(r) \sin \theta
\biggl[1 - {{\rm D}_r \over 2} \biggl(1 - {4  \over 5} (2q+1) {\sigma_c \over 1 + \sigma_c}\biggr) \biggr]
\nonumber\\
&& \times \nabla_r \ln \left(\meanrho^{\, \mu_{\rm v}} \, u_{\rm rms} \right)  ,
\label{CWB24}
\end{eqnarray}
where
\begin{eqnarray}
\mu_{\rm v} &=& \biggl\{1 - {\sigma_c \over 2(1 + \sigma_c)} -{2{\rm D}_r \over 5} \biggl[
\Big(3 q + {31 \over 8}\Big) {\sigma_c \over 1 + \sigma_c}
\nonumber\\
&& - 4 -2q \biggr] \biggr\} \,  \biggl[1 - {{\rm D}_r \over 2} \biggl(1 - {4  \over 5} (2q+1)
{\sigma_c \over 1 + \sigma_c}\biggr) \biggr]^{-1} .
\nonumber\\
\label{CWB21}
\end{eqnarray}
To derive equations~(\ref{CWB20}) and~(\ref{CWB24}), we use equations~(\ref{WB1}) and~(\ref{WB4}).
The parameter ${\rm D}_r$ characterising the radial differential rotation, is defined as
\begin{eqnarray}
{\rm D}_r = {\partial \ln \delta\Omega \over \partial  \ln r} .
\label{WB25}
\end{eqnarray}

The joint action of the $\alpha$ effect and the large-scale shear causes the dynamo
resulting in the generation of the large-scale magnetic field.
For illustration, we consider the axisymmetric mean-field $\alpha^2 \, \Omega$ dynamo, so that
the large-scale magnetic field can be written as  $ \meanBB = \meanB_\varphi
{\bm e}_{\varphi} + \bec{\nabla} {\bf \times} (\meanA {\bm e}_{\varphi})$.
For simplicity, we study the mean-field dynamo in a thin shell,
neglecting the curvature of the shell and replace it by a flat slab.
We consider a kinematic dynamo problem,
assuming for simplicity that the kinetic $\alpha$ effect
is a constant.
The mean-field dynamo equations in a dimensionless form are given by:
\begin{eqnarray}
{\partial \meanB_\varphi \over \partial t} &=& \left[R_\alpha \, R_\omega \, \sin \theta {\partial \over \partial \theta}
- R_\alpha^2 \,  \left({\partial^2 \over \partial \theta^2} - \kappa^2 \right)
 \right]\meanA
\nonumber\\
&& + \left({\partial^2 \over \partial \theta^2} - \kappa^2 \right)\meanB_\varphi ,
\label{TM1}
\end{eqnarray}
\begin{eqnarray}
{\partial \meanA \over \partial t} &=& \alpha \meanB_\varphi + \left({\partial^2
\over \partial \theta^2} - \kappa^2 \right) \meanA .
\label{TM2}
\end{eqnarray}
where $\alpha \equiv \alpha_{\varphi\varphi}$, for simplicity we average the dynamo equations over $r$ and use the no-$r$ model.
In particular, the terms describing turbulent diffusion of the mean magnetic field
in the radial direction in equations~(\ref{TM1}) and~(\ref{TM2}) in the framework
of the no-$r$ model are given as $-\kappa^2 \meanB_\varphi$
and $-\kappa^2\meanA$ \citep{KSR16}, where
the parameter $\kappa$ is determined by the following equation:
$\int_{r_{\rm c}}^{1} (\partial^2 \meanB_\varphi / \partial r^2) \,dr = - (\kappa^2/3) \meanB_\varphi$.
Here the radius $r$ varies from $r_{\rm c}$ to $1$ inside the convective shell.

Equations~(\ref{TM1})--(\ref{TM2}) are written in dimensionless variables:
the coordinate $r$ is measured in the units of the radius $R_\ast$, the time $t$
is measured in the units of turbulent magnetic diffusion time $R_\ast^2 / \eta_{_{T}}$,
and the toroidal component $\meanB_\varphi(t,\theta)$ of the mean magnetic field is
measured in the units of $\meanB_{\rm eq}= u_0 \, \sqrt{\mu_0 \meanrho_\ast}$.
The  magnetic potential $\meanA(t,\theta)$ of the poloidal field is measured in the units of
$R_{\alpha} R_\ast \meanB_{\rm eq}$, where
\begin{eqnarray}
R_{\alpha} = {\alpha_\ast \, R_\ast \over \eta_{_{T}}} = {\ell_0^2 \, \delta\Omega \, R_\ast \over H_\rho \, \eta_{_{T}} } ,
\label{TMM3}
\end{eqnarray}
the fluid density $\meanrho$ is measured in the units $\meanrho_\ast$,
the differential rotation $\delta\Omega$ is measured in units of the maximal value
of the angular velocity $\Omega$,
the $\alpha$ effect is measured in units of the maximum value of the
kinetic $ \alpha $ effect $\alpha_\ast$,
the integral scale of the turbulent motions
$\ell_0$ and the characteristic turbulent velocity $u_0$ at the scale $\ell_0$ are measured in units of their
maximum values in the turbulent region, and
the turbulent magnetic diffusion coefficient is $\eta_{_{T}}=\ell_0 \, u_0 / 3$.
The dynamo number is defined as $D = R_\alpha
R_\omega$, where $R_\omega = \delta \Omega \, R_\ast^2 / \eta_{_{T}}$.

Equations~(\ref{TM1}) and~(\ref{TM2})
describe the dynamo waves propagating from the central
latitudes towards the equator when the dynamo number is negative.
We seek a solution for equations~(\ref{TM1})--(\ref{TM2}) as a real part of the following functions:
$\overline{A}=A_0 \exp(\tilde \gamma t - {\rm i} \, K \, \theta)$ and
$\overline{B}_\varphi = B_0 \exp(\tilde \gamma t - {\rm i} \, K \, \theta)$,
where $\tilde \gamma=\gamma + {\rm i} \, \omega$.
The growth rate of the dynamo instability and the frequency of the dynamo waves are given by
\citep{KRS23}:
\begin{eqnarray}
\gamma &=& {R_{\alpha} R_{\alpha}^{\rm cr} \over \sqrt{2}} \left[\left[1 + \left({\zeta R_\omega \over R_{\alpha} R_{\alpha}^{\rm cr}}\right)^2\right]^{1/2} + 1 \right]^{1/2}
- \left(R_{\alpha}^{\rm cr}\right)^2 ,
\nonumber\\
\label{TM5}
\end{eqnarray}
\begin{eqnarray}
\omega = - {\rm sgn}(R_\omega) \, {R_{\alpha} R_{\alpha}^{\rm cr} \over \sqrt{2}} \left[\left[1 + \left({\zeta R_\omega \over R_{\alpha} R_{\alpha}^{\rm cr}}\right)^2\right]^{1/2} - 1 \right]^{1/2}  ,
\label{TM6}
\end{eqnarray}
where $\zeta^2=1 - \left(\kappa/R_{\alpha}^{\rm cr}\right)^2$.
Here we took into account that
$(x + {\rm i}y)^{1/2}= \pm (X + {\rm i}Y)$, where
$X = 2^{-1/2} \, [(x^2+y^2)^{1/2} + x]^{1/2}$ and $Y = {\rm sgn}(y) \, 2^{-1/2} \, [(x^2+y^2)^{1/2} - x]^{1/2}$.
The threshold $R_{\alpha}^{\rm cr}$ for the mean-field dynamo instability, defined by the conditions
$\gamma=0$ and $R_\omega=0$, is given by $R_{\alpha}^{\rm cr}=(K^2 + \kappa^2)^{1/2}$.
The energy ratio of poloidal $\overline{B}_{\rm pol} =R_{\alpha} R_{\alpha}^{\rm cr} \, \overline{A}$ and toroidal $\overline{B}_\varphi$ mean magnetic field components are given by
\begin{eqnarray}
{\meanB_{\rm pol}^2 \over \meanB_\varphi^2} =  \left[1 + \left({\zeta R_\omega \over R_{\alpha} R_{\alpha}^{\rm cr}}\right)^2\right]^{-1/2} ,
\label{TM12}
\end{eqnarray}
and the phase shift $\delta$ between the toroidal field $\overline{B}_\varphi$ and the magnetic vector potential $\overline{A}$ is
\begin{eqnarray}
\sin(2\delta) =  - \zeta R_\omega \,  \left[\left(R_{\alpha} R_{\alpha}^{\rm cr} \right)^{2} + \zeta^2 R_\omega^2\right]^{-1/2} .
\label{TM14}
\end{eqnarray}

Now we apply the developed theory to the various astrophysical turbulent clouds.
Let us first discuss a scenario of formation the large-scale shear
motions in colliding protogalactic clouds \citep[see, e.g.,][]{CH91,CH93,WBL98,BWL02,RKCL06}.
Jean's process of gravitational instability and fragmentation can cause a very clumsy
state of cosmic matter at the epoch of galaxy formation. A complex
system of rapidly moving gaseous fragments embedded into rare gas
might appear in some regions of protogalactic matter. Supersonic
contact collisions of these protogalactic clouds might play a role
of an important elementary process in a complex nonlinear dynamics
of protogalactic medium. The supersonic contact non-central
collisions of these protogalactic clouds could lead to their
coalescence, formation of large-scale shear motions and transformation of their
initial orbital momentum into the spin momentum of the merged
condensations bound by its condensations \citep{CH93}.

Two-dimensional hydrodynamical models for inelastic non-central
cloud-cloud collisions in the protogalactic medium have been
developed by \cite{CH93}. An evolutionary picture of the
collision is as follows. At the first stage of the process the
standard dynamical structure, i.e., two shock fronts and tangential
discontinuity between them arise in the collision zone. Compression
and heating of gas which crosses the shock fronts occurs. The
heating entails intensive radiation emission and considerable energy
loss by the system which promotes gravitational binding of the cloud
material.

At the second stage of the process a dense core forms at
the central part of the clump. In the vicinity of the core two kinds
of jets form: "flyaway" jets of the material (which does not undergo
the direct contact collision) and internal jets sliding along the
curved surface of the tangential discontinuity. The flyaway jets are
subsequently torn off, having overcome the gravitational attraction
of the clump whereas the internal jets remain bound in the clump.
When the shock fronts reach the outer boundaries of the clump, the
third stage of the process starts. Shocks are replaced by the
rarefaction waves and overall differential rotation and large-scale
shear motions arise. This structure can be considered as a model of
the protogalactic condensation \citep{CH93}.

\begin{table}
\label{tab1}
\begin{tabular}{|l|c|c|}
\multicolumn{3}{c}{Table~\ref{tab1}}\\
\multicolumn{3}{c}{The parameters of clouds}\\
\hline
                              & PGC     & PSC
 \\
 \hline
   &   & \\
 Mass  & $M \leq 10^{10} \, M_\odot$  & $M \leq M_\odot$ \\
 \hline
   &   & \\
 $R \, $ (cm)  & $10^{23}$  & $10^{17}$\\
\hline
   &   & \\
 $\meanU \, $ (cm/s)  & $10^{6} - 10^{7}$  & $10^{5} - 10^{6}$
 \\
\hline
   &   & \\
 $\meanrho \, \,$ (g/cm$^{3}$)        & $10^{-26}$   & $(1-5) \times 10^{-19}$
 \\
 \hline
   &   & \\
 $\Delta \meanU \, $ (cm/s)  & $10^{6} - 10^{7}$  & $10^{5}$
 \\
 \hline
   &   & \\
 $\Delta R \, $ (cm)      & $2 \times 10^{23}$ & $10^{16} - 10^{17}$
 \\
 \hline
   &   & \\
 $S \,$ ($s^{-1}$)        & $(0.5 - 5)  \times 10^{-16}$ & $10^{-12} - 10^{-11}$
 \\
\hline
   &   & \\
 $u_0 \, $ (cm/s)    & $10^{6} - 10^{7}$ & $10^{4}$
 \\
\hline
   &   & \\
 $\ell_0 \, $ (cm)  & $10^{22}$ & $10^{15} - 10^{16}$
 \\
\hline
   &   & \\
 $\tau_0  \, $ (years)  & $(0.3 - 3) \times 10^{8}$ & $(0.3 - 3) \times 10^{4}$
 \\
\hline
   &   & \\
 $\eta_{_{T}} \,$ (cm$^2$/s)  & $(0.3 - 3) \times 10^{28}$
 & $(0.3 - 3) \times 10^{19}$
 \\
\hline
   &   & \\
 $t_\eta \, $ (years) & $(0.3 - 3) \times 10^{9}$
 & $10^{6} - 10^{7}$
 \\
\hline
   &   & \\
 $\alpha \, $ (cm/s)  & $10^{4} - 10^{5}$  & $10^{2} - 10^{4}$
 \\
\hline
   &   & \\
 $V^{\rm eff}_\varphi  \, $ (cm/s)  & $10^{3} - 10^{4}$  & $10^{2} - 10^{4}$
 \\
\hline
\end{tabular}
\end{table}

The formed large-scale sheared motions are superimposed on small-scale
turbulence.
There are two important characteristics of the protogalactic cloud -
cloud collisions: the mass bound in the resulting clump and the spin
momentum acquired by it. These characteristics depend on the
relative velocity and impact parameter of the collision  \citep{CH93}.

The parameters of protogalactic clouds are as follows \citep[see, e.g.,][]{CH91,CH93,WBL98,BWL02,RKCL06}: the
mass is $M \leq 10^{10} \, M_\odot$, the radius is $R \sim 10^{23}$
cm, the internal temperature is $\meanT \sim 10^4 $ K, the mean radial velocity
of the cloud is $\meanU  \sim 10^{6}$--$10^{7}$ cm/s, where $M_\odot$ is
the solar mass.
Some other parameters for the protogalactic clouds
(PGC) are given in Table~\ref{tab1}.

We use the following notations:
$\Delta R$ is the characteristic scale of the
mean velocity inhomogeneity,
$\Delta\meanU$ is the typical velocity change across $\Delta R$,
$S = \Delta \meanU  / \Delta R$ is the mean
velocity shear, $u_0$ is the characteristic turbulent velocity,
$\ell_0$ is the integral scale of turbulent motions, $\tau_0 = \ell_0 /
u_0$ is the characteristic turbulent time, $\eta_{_{T}}$ is the
turbulent magnetic diffusivity, and $t_\eta = (\Delta R)^2 /
\eta_{_{T}}$ is the turbulent diffusion time.
Using the parameters given in Table~\ref{tab1}, we
estimate the $\alpha$ effect as $\alpha \equiv |\alpha_{\varphi\varphi}| \sim 10^4$--$10^5$ cm/s
and the effective pumping velocity of the mean magnetic field is estimated as
$|V^{\rm eff}_\varphi | \sim 10^3$--$10^4$ cm/s.
The $\alpha$ effect in combination  with the large-scale shear motions can cause generation
of large-scale magnetic field.

An important feature of the dynamics of the interstellar matter is
fairly rapid motions of relatively dense matter fragments
(protostellar clouds) embedded in to rare gas. The origin of
protostellar clouds might be a result of fragmentation of the core
of large molecular clouds. Supersonic and inelastic collisions of
the protostellar clouds can cause merging of the clouds and
formation of a condensation. A non-central collision of the
protostellar clouds can cause conversion of initial orbital momentum
of the clouds in to spin momentum and formation of differential
rotation and shear motions \citep{CH91}.

The internal part of the
condensation would have only slow rotation because the initial
matter motions could be almost stopped in the zone of direct cloud
contact. On the other hand, the minor outer part of the merged cloud
matter of the condensation would have very rapid rotation due to the
initial motions of that portions of cloud materials which would not
stop in this zone because they do not undergo any direct cloud
collision \citep{CH91}.
This material could keep its motion on
gravitationally bound orbits around the major internal body
condensation. The formed large-scale sheared motions are
superimposed on small-scale interstellar turbulence.

In the supersonic and inelastic collision of the protostellar clouds,
an essential part of the initial kinetic energy
that is lost during the mass lost, is due to dissipation and
subsequent radiative emission. The cooling time scale for the
material compressed in the collision would be less than the time
scale of the hydrodynamic processes.

The parameters of protostellar clouds
are as follows \citep[see, e.g.,][]{CH91,RKCL06}: a mass is $M \leq M_\odot$, the radius is $R \sim
10^{17}$ cm, the internal temperature is $\meanT \sim 10$ K, the mean radial
velocity of the cloud is $\meanU \sim 10^{5}$--$10^{6} $ cm/s. Some other
parameters for the protostellar clouds (PSC) are given in Table~\ref{tab1}.
Using the parameters given in Table~\ref{tab1}, we
find the $\alpha$ effect as $\alpha \equiv |\alpha_{\varphi\varphi}|  \sim 10^2$--$10^4$ cm/s
and the effective pumping velocity of the mean magnetic field is
$|V^{\rm eff}_\varphi | \sim 10^2$--$10^4$ cm/s.

\section{Conclusions}
\label{sect-7}

In the present study we determine the $\alpha$ effect and the effective  pumping velocity
of a large-scale magnetic field caused by
a combined effect of the density-stratified turbulence and
a large-scale shear for arbitrary Mach numbers.
These phenomena are derived applying the spectral $\tau$ approach
that is valid for large fluid and magnetic Reynolds numbers.

We demonstrate that the finite Mach number effects
does not affect the contributions caused by the inhomogeneity of turbulence
to the $\alpha$ tensor, but they influence the effective pumping velocity of the mean magnetic field.
In addition, the isotropic part of the ${\bm \alpha}$ tensor is independent of the exponent
of the turbulent kinetic energy spectrum for this system.
On the other hand, the anisotropic part of the ${\bm \alpha}$ tensor
depends on this exponent, and the latitudinal profile of differential rotation also contributes
to this anisotropic part of the ${\bm \alpha}$ tensor.
The latter may be important for the dynamo operation in the upper parts of the solar and stellar convection zones.
We also find an additional contribution to the effective pumping velocity of the mean magnetic field
that is proportional to the product of the fluid density stratification and the divergence of the mean fluid velocity
caused by collapsing (or expanding) astrophysical clouds.
On the other hand, we show that the ${\bm \alpha}$ tensor is independent of the effects of collapsing
(or expanding) clouds.

These effects may be the reasons for generation of the large-scale magnetic field in
protoplanetary discs, colliding protogalactic clouds,
merging protostellar clouds, solar and stellar convective zones.
In particular, the theoretical results of Section~\ref{subsect-6.2}
are directly applicable to the solar and stellar convective zones
(which is a subject of a separate study).

\section*{Acknowledgments}

We thank M. Rheinhardt for his suggestions which have
significantly improved the paper.
IR would like to thank the Isaac Newton Institute for
Mathematical Sciences, Cambridge, for support and hospitality
during the program ‘Anti-diffusive dynamics: from sub-cellular to
astrophysical scales’ (April - June 2024), where this work was initiated.
IR also acknowledges the discussions with some participants of the Nordita
Scientific Programs on ‘Stellar convection: modelling, theory and
observations’ (September 2024)
and ‘Numerical Simulations of Early Universe Sources of Gravitational Waves’
(July - August 2025), Stockholm.
IR would like to thank the Nordita for support and hospitality during the programs.

\section*{Data Availability}

There are no new data associated with this article.

\noindent

\bibliographystyle{mnras}
\bibliography{accepted-paper-MNRAS}

\begin{thebibliography}{}
\makeatletter
\relax
\def\mn@urlcharsother{\let\do\@makeother \do\$\do\&\do\#\do\^\do\_\do\%\do\~}
\def\mn@doi{\begingroup\mn@urlcharsother \@ifnextchar [ {\mn@doi@}
  {\mn@doi@[]}}
\def\mn@doi@[#1]#2{\def\@tempa{#1}\ifx\@tempa\@empty \href
  {http://dx.doi.org/#2} {doi:#2}\else \href {http://dx.doi.org/#2} {#1}\fi
  \endgroup}
\def\mn@eprint#1#2{\mn@eprint@#1:#2::\@nil}
\def\mn@eprint@arXiv#1{\href {http://arxiv.org/abs/#1} {{\tt arXiv:#1}}}
\def\mn@eprint@dblp#1{\href {http://dblp.uni-trier.de/rec/bibtex/#1.xml}
  {dblp:#1}}
\def\mn@eprint@#1:#2:#3:#4\@nil{\def\@tempa {#1}\def\@tempb {#2}\def\@tempc
  {#3}\ifx \@tempc \@empty \let \@tempc \@tempb \let \@tempb \@tempa \fi \ifx
  \@tempb \@empty \def\@tempb {arXiv}\fi \@ifundefined
  {mn@eprint@\@tempb}{\@tempb:\@tempc}{\expandafter \expandafter \csname
  mn@eprint@\@tempb\endcsname \expandafter{\@tempc}}}

\bibitem[\protect\citeauthoryear{Birk, Wiechen  \& Lesch}{Birk
  et~al.}{2002}]{BWL02}
Birk G.,  Wiechen H.,   Lesch H.,  2002, A\&A, 393, 685

\bibitem[\protect\citeauthoryear{Brandenburg \& Ntormousi}{Brandenburg \&
  Ntormousi}{2025}]{BN25}
Brandenburg A.,  Ntormousi E.,  2025, ApJ, 990, 223

\bibitem[\protect\citeauthoryear{Brandenburg \& Subramanian}{Brandenburg \&
  Subramanian}{2005a}]{Brandenburg(2005)}
Brandenburg A.,  Subramanian K.,  2005a, Phys. Rep., 417, 1

\bibitem[\protect\citeauthoryear{Brandenburg \& Subramanian}{Brandenburg \&
  Subramanian}{2005b}]{BS05b}
Brandenburg A.,  Subramanian K.,  2005b, A\&A, 439, 835

\bibitem[\protect\citeauthoryear{Brandenburg, R{\"a}dler  \& Kemel}{Brandenburg
  et~al.}{2012}]{BRK12b}
Brandenburg A.,  R{\"a}dler K.-H.,   Kemel K.,  2012, A\&A, 539, A35

\bibitem[\protect\citeauthoryear{Brandenburg, Gressel, K{\"a}pyl{\"a},
  Kleeorin, Mantere  \& Rogachevskii}{Brandenburg et~al.}{2013}]{BGKR13}
Brandenburg A.,  Gressel O.,  K{\"a}pyl{\"a} P.~J.,  Kleeorin N.,  Mantere
  M.~J.,   Rogachevskii I.,  2013, ApJ, 762, 127

\bibitem[\protect\citeauthoryear{Brandenburg, Rogachevskii  \&
  Kleeorin}{Brandenburg et~al.}{2016}]{BKR16}
Brandenburg A.,  Rogachevskii I.,   Kleeorin N.,  2016, New J. Physics, 18,
  125011

\bibitem[\protect\citeauthoryear{Brandenburg, Rogachevskii  \&
  Schober}{Brandenburg et~al.}{2023}]{BRS23}
Brandenburg A.,  Rogachevskii I.,   Schober J.,  2023, MNRAS, 518, 6367

\bibitem[\protect\citeauthoryear{Chernin}{Chernin}{1991}]{CH91}
Chernin A.~D.,  1991, Astrophys. Space Science, 186, 159

\bibitem[\protect\citeauthoryear{Chernin}{Chernin}{1993}]{CH93}
Chernin A.~D.,  1993, A\&A, 267, 315

\bibitem[\protect\citeauthoryear{Elperin, Kleeorin  \& Rogachevskii}{Elperin
  et~al.}{1998}]{EKR98}
Elperin T.,  Kleeorin N.,   Rogachevskii I.,  1998, Phys. Rev. Lett., 81, 2898

\bibitem[\protect\citeauthoryear{Hodgson \& Brandenburg}{Hodgson \&
  Brandenburg}{1998}]{HB98}
Hodgson L.~S.,  Brandenburg A.,  1998, A\&A, 330, 1169

\bibitem[\protect\citeauthoryear{Hopkins}{Hopkins}{2016a}]{HOP16}
Hopkins P.~F.,  2016a, MNRAS, 455, 89

\bibitem[\protect\citeauthoryear{Hopkins}{Hopkins}{2016b}]{HOP16b}
Hopkins P.~F.,  2016b, MNRAS, 456, 2383

\bibitem[\protect\citeauthoryear{Hubbard}{Hubbard}{2016}]{HUB16}
Hubbard A.,  2016, MNRAS, 456, 3079

\bibitem[\protect\citeauthoryear{Irshad, P, K  \& A}{Irshad
  et~al.}{2025}]{IBSS25}
Irshad P B.,  K S.,   A S.,  2025, arXiv:2503.19131

\bibitem[\protect\citeauthoryear{Kleeorin \& Rogachevskii}{Kleeorin \&
  Rogachevskii}{2022}]{KR22}
Kleeorin N.,  Rogachevskii I.,  2022, MNRAS, 515, 5437

\bibitem[\protect\citeauthoryear{Kleeorin \& Rogachevskii}{Kleeorin \&
  Rogachevskii}{2025}]{KR25}
Kleeorin N.,  Rogachevskii I.,  2025, Phys. Fluids, 37, 065152

\bibitem[\protect\citeauthoryear{Kleeorin, Safiullin, Kleeorin, Porshnev,
  Rogachevskii  \& Sokoloff}{Kleeorin et~al.}{2016}]{KSR16}
Kleeorin Y.,  Safiullin N.,  Kleeorin N.,  Porshnev S.,  Rogachevskii I.,
  Sokoloff D.,  2016, MNRAS, 460, 3960

\bibitem[\protect\citeauthoryear{Kleeorin, Rogachevskii, Safiullin, Gershberg
  \& Porshnev}{Kleeorin et~al.}{2023}]{KRS23}
Kleeorin N.,  Rogachevskii I.,  Safiullin N.,  Gershberg R.,   Porshnev S.,
  2023, Monthly Notices of the Royal Astronomical Society, 526, 1601

\bibitem[\protect\citeauthoryear{Krause \& R{\"a}dler}{Krause \&
  R{\"a}dler}{1980}]{Krause(1980)}
Krause F.,  R{\"a}dler K.-H.,  1980, Mean-Field Magnetohydrodynamics and Dynamo
  Theory.
Oxford: Pergamon Press

\bibitem[\protect\citeauthoryear{Moffatt}{Moffatt}{1978}]{Moffatt(1978)}
Moffatt H.~K.,  1978, Magnetic Field Generation in Electrically Conducting
  Fluids.
Cambridge: Cambridge University Press

\bibitem[\protect\citeauthoryear{Moffatt \& Dormy}{Moffatt \&
  Dormy}{2019}]{MD2019}
Moffatt H.~K.,  Dormy E.,  2019, Self-Exciting Fluid Dynamos.
Cambridge: Cambridge University Press

\bibitem[\protect\citeauthoryear{Orszag}{Orszag}{1970}]{O70}
Orszag S.~A.,  1970, J. Fluid Mech., 41, 363

\bibitem[\protect\citeauthoryear{Pan, Padoan, Scalo, Kritsuk  \& Norman}{Pan
  et~al.}{2011}]{PPK11}
Pan L.,  Padoan P.,  Scalo J.,  Kritsuk A.~G.,   Norman M.~L.,  2011, ApJ, 740,
  6

\bibitem[\protect\citeauthoryear{Parker}{Parker}{1979}]{Parker(1979)}
Parker E.~N.,  1979, Cosmical Magnetic Fields: Their Origin and their Activity.
Oxford: Clarendon Press

\bibitem[\protect\citeauthoryear{Pouquet, Frisch  \& L{\'e}orat}{Pouquet
  et~al.}{1976}]{PFL76}
Pouquet A.,  Frisch U.,   L{\'e}orat J.,  1976, J. Fluid Mech., 77, 321

\bibitem[\protect\citeauthoryear{R{\"a}dler \& Stepanov}{R{\"a}dler \&
  Stepanov}{2006}]{RS06}
R{\"a}dler K.-H.,  Stepanov R.,  2006, Phys. Rev. E, 73, 056311

\bibitem[\protect\citeauthoryear{Roberts \& Soward}{Roberts \&
  Soward}{1975}]{RS75}
Roberts P.~H.,  Soward A.~M.,  1975, Astron. Nachr., 296, 49

\bibitem[\protect\citeauthoryear{Rogachevskii}{Rogachevskii}{2021}]{RI21}
Rogachevskii I.,  2021, Introduction to Turbulent Transport of Particles,
  Temperature and Magnetic Fields.
Cambridge: Cambridge University Press

\bibitem[\protect\citeauthoryear{Rogachevskii \& Kleeorin}{Rogachevskii \&
  Kleeorin}{2003}]{RK03}
Rogachevskii I.,  Kleeorin N.,  2003, Phys. Rev. E, 68, 036301

\bibitem[\protect\citeauthoryear{Rogachevskii \& Kleeorin}{Rogachevskii \&
  Kleeorin}{2004}]{RK04}
Rogachevskii I.,  Kleeorin N.,  2004, Phys. Rev. E, 70, 046310

\bibitem[\protect\citeauthoryear{Rogachevskii \& Kleeorin}{Rogachevskii \&
  Kleeorin}{2021a}]{RK21B}
Rogachevskii I.,  Kleeorin N.,  2021a, Phys. Rev. E, 103, 013107

\bibitem[\protect\citeauthoryear{Rogachevskii \& Kleeorin}{Rogachevskii \&
  Kleeorin}{2021b}]{RK21A}
Rogachevskii I.,  Kleeorin N.,  2021b, MNRAS, 508, 1296

\bibitem[\protect\citeauthoryear{Rogachevskii, Kleeorin, Chernin  \&
  Liverts}{Rogachevskii et~al.}{2006}]{RKCL06}
Rogachevskii I.,  Kleeorin N.,  Chernin A.~D.,   Liverts E.,  2006, Astron.
  Nachr., 327, 591

\bibitem[\protect\citeauthoryear{Rogachevskii, Kleeorin, K{\"a}pyl{\"a}  \&
  Brandenburg}{Rogachevskii et~al.}{2011}]{RKK11}
Rogachevskii I.,  Kleeorin N.,  K{\"a}pyl{\"a} P.~J.,   Brandenburg A.,  2011,
  Phys. Rev. E, 84, 056314

\bibitem[\protect\citeauthoryear{Rogachevskii, Kleeorin, Brandenburg  \&
  Eichler}{Rogachevskii et~al.}{2012}]{RKB12}
Rogachevskii I.,  Kleeorin N.,  Brandenburg A.,   Eichler D.,  2012, ApJ, 753,
  6

\bibitem[\protect\citeauthoryear{Rogachevskii, Ruchayskiy, Boyarsky,
  Fr{\"o}hlich, Kleeorin, Brandenburg  \& Schober}{Rogachevskii
  et~al.}{2017}]{RRB17}
Rogachevskii I.,  Ruchayskiy O.,  Boyarsky A.,  Fr{\"o}hlich J.,  Kleeorin N.,
  Brandenburg A.,   Schober J.,  2017, ApJ, 846, 153

\bibitem[\protect\citeauthoryear{Rogachevskii, Kleeorin  \&
  Brandenburg}{Rogachevskii et~al.}{2018}]{RKB18}
Rogachevskii I.,  Kleeorin N.,   Brandenburg A.,  2018, J. Plasma Phys., 84,
  735840502

\bibitem[\protect\citeauthoryear{R{\"u}diger \& Kichatinov}{R{\"u}diger \&
  Kichatinov}{1993}]{RUK93}
R{\"u}diger G.,  Kichatinov L.~L.,  1993, A\&A, 269, 581

\bibitem[\protect\citeauthoryear{R{\"u}diger, Hollerbach  \&
  Kitchatinov}{R{\"u}diger et~al.}{2013}]{Ruediger(2013)}
R{\"u}diger G.,  Hollerbach R.,   Kitchatinov L.~L.,  2013, Magnetic Processes
  in Astrophysics: Theory, Simulations, Experiments.
Weinheim: John Wiley \& Sons

\bibitem[\protect\citeauthoryear{Ruzmaikin, Shukurov  \& Sokoloff}{Ruzmaikin
  et~al.}{1988}]{Ruzmaikin(1988)}
Ruzmaikin A.,  Shukurov A.~M.,   Sokoloff D.~D.,  1988, Magnetic Fields of
  Galaxies.
Dordrecht: Kluwer Academic

\bibitem[\protect\citeauthoryear{Schober, Rogachevskii, Brandenburg, Boyarsky,
  Fr{\"o}hlich, Ruchayskiy  \& Kleeorin}{Schober et~al.}{2018}]{SRB18}
Schober J.,  Rogachevskii I.,  Brandenburg A.,  Boyarsky A.,  Fr{\"o}hlich J.,
  Ruchayskiy O.,   Kleeorin N.,  2018, ApJ, 858, 124

\bibitem[\protect\citeauthoryear{Shukurov \& Subramanian}{Shukurov \&
  Subramanian}{2021}]{SS21}
Shukurov A.,  Subramanian K.,  2021, Astrophysical Magnetic Fields: From
  Galaxies to the Early Universe.
Cambridge University Press

\bibitem[\protect\citeauthoryear{Steenbeck, Krause  \& R{\"a}dler}{Steenbeck
  et~al.}{1966}]{Steenbeck(1966)}
Steenbeck M.,  Krause F.,   R{\"a}dler K.-H.,  1966, Zeitschrift f{\"u}r
  Naturforschung A, 21, 369

\bibitem[\protect\citeauthoryear{Wiechen, Birk  \& Lesch}{Wiechen
  et~al.}{1998}]{WBL98}
Wiechen H.,  Birk G.,   Lesch H.,  1998, A\&A, 334, 388

\bibitem[\protect\citeauthoryear{Zeldovich, Ruzmaikin  \& Sokoloff}{Zeldovich
  et~al.}{1983}]{Zeldovich(1983)}
Zeldovich Y.~B.,  Ruzmaikin A.~A.,   Sokoloff D.~D.,  1983, Magnetic Fields in
  Astrophysics.
New-York: Gordon and Breach

\makeatother
\end{thebibliography}

\appendix

\section{Identities used for derivation of EMF}
\label{Appendix A}

The tensors $I_{ijmn}(\meanUU)$, $J_{ijmn}(\meanUU)$ and $E_{ijmn}(\meanUU)$
in equations~(\ref{B3})--(\ref{NNB4}) are given by \citep{KR22}:
\begin{eqnarray}
&& I_{ijmn}(\meanUU) = \biggl\{2 k_{iq} \delta_{mp}
\delta_{jn} + 2 k_{jq} \delta_{im} \delta_{pn} - \delta_{im}
\delta_{jq} \delta_{np}
\nonumber\\
 && \quad - \delta_{iq} \delta_{jn} \delta_{mp}  + 4 k_{pq} \delta_{im} \delta_{jn}  + \delta_{im} \delta_{jn}
k_{q} {\partial \over \partial k_{p}}
\nonumber\\
 &&\quad - {{\rm  i} \, \lambda_r \over 2 k^2} \, \biggl[\Big(k_i  \delta_{jn} \delta_{pm}
- k_j \delta_{im} \delta_{pn}\Big) \, \Big(2 k_{rq}  - \delta_{rq}\Big)
\nonumber\\
 && \quad + k_q \Big(\delta_{ip} \delta_{jn} \delta_{rm} - \delta_{im} \delta_{jp} \delta_{rn}\Big)
- 2 k_{pq}\Big(k_i  \delta_{jn} \delta_{rm}
\nonumber\\
 && \quad - k_j \delta_{im} \delta_{rn}\Big)
\biggr] \biggr\} \nabla_{p} \meanU_{q},
\label{B5}
\end{eqnarray}
\begin{eqnarray}
&& J_{ijmn}(\meanUU) = \biggl\{2 k_{iq} \delta_{jn}
\delta_{pm} - \delta_{iq} \delta_{jn} \delta_{pm} + \delta_{im} \delta_{jq} \delta_{pn}
\nonumber\\
 && \quad + 2 k_{pq} \delta_{im} \delta_{jn} + \delta_{im} \delta_{jn} k_{q} {\partial \over \partial k_{p}}
  - {{\rm  i} \, \lambda_r \over 2 k^2} \, \biggl[
k_i  \delta_{jn} \delta_{pm}
\nonumber\\
&& \quad \times\Big(2 k_{rq} - \delta_{rq}\Big)
+ \delta_{jn} \delta_{rm} \, \Big(k_q   \, \delta_{ip} - 2 k_i  \, k_{pq} \Big)
\biggr] \biggr\} \nabla_{p} \meanU_{q} ,
\nonumber\\
\label{B6}
\end{eqnarray}
\begin{eqnarray}
&& E_{ijmn}(\meanUU) = \biggl[\delta_{im} \delta_{jq} \delta_{pn}+
 \delta_{iq} \delta_{jn} \delta_{pm}
 \nonumber\\
 && \qquad
 + \delta_{im} \delta_{jn}
k_{q} {\partial \over \partial k_{p}} \biggr] \, \nabla_{p} \meanU_{q} .
\label{NNB5}
\end{eqnarray}
Equations~(\ref{B5})--(\ref{NNB5}) are valid for weak large-scale shear ($\meanW \tau_0 \ll 1$),
where $\meanWW$ is the mean vorticity, and
we neglected the second-order derivatives of the mean velocity $\meanUU$.
The reason of the appearance of the stratification parameter $\lambda_i$
in the tensors $I_{ijmn}$ and $J_{ijmn}$ is caused
by the exclusion of the gradient of pressure fluctuations from the Navier-Stokes
equation~(\ref{B0}) by taking twice curl from this equation.
On the other hand, the parameter $\Lambda_i$ that characterises the inhomogeneity of turbulence
cannot enter in the tensors $I_{ijmn}$ and $J_{ijmn}$. It appears only
in the tensor $f_{ij}^{(0)}$.

To derive expression for the contributions to the turbulent electromotive force
caused by a density-stratified and inhomogeneous turbulence with a non-uniform large-scale flow
and low Mach numbers, we use the following identities:
\begin{eqnarray}
&& {\cal E}^{(1a)}_f  = \left\langle {\bm u}^2 \right\rangle^{(0)} \, {\meanB_s \varepsilon_{fij} \over 8 \pi}
\int \tau^2(k) \,  k_s \,  I_{ijmn} \,  E(k) \, \big(\tilde \lambda_m k_{n}
\nonumber\\
&& \quad - \tilde \lambda_n k_{m}\big) \, k^{-4}\,  d{\bm k}
= {4 \over 45} \ell_0^2  \, \meanB_j  \tilde \lambda_r  \Big[
(2q-1)
\varepsilon_{frn} \, \Delta_{pqjn}
\nonumber\\
&& \quad + 5  \Big(\varepsilon_{fjq} \,  \delta_{rp}
+\varepsilon_{frp} \,  \delta_{qj}
+ \varepsilon_{fqr} \,  \delta_{jp} \Big) \Big] \nabla_p \meanU_{q} ,
\label{D1}
\end{eqnarray}

\begin{eqnarray}
&& {\cal E}^{(1b)}_f  = \left\langle {\bm u}^2 \right\rangle^{(0)} \, {\meanB_s \varepsilon_{fij} \over 8 \pi}
\int \tau^2(k)  \, k_s  \, J_{ijmn}  \, E(k) \, \big(\tilde \lambda_m k_{n}
\nonumber\\
&& \quad  - \tilde \lambda_n k_{m}\big) \,  k^{-4}  \, d{\bm k}
= {4 \over 45} \,   \ell_0^2  \, \meanB_j  \,  \tilde \lambda_r
\, \Big[2(q + 3)\, \varepsilon_{frn} \, \Delta_{pqjn}
\nonumber\\
&& \quad+ 5\varepsilon_{fpr}
\,  \delta_{qj} \Big] \nabla_p \meanU_{q} ,
\label{D2}
\end{eqnarray}
where $\tilde \lambda_i = \lambda_i - \Lambda_i/2$, and
\begin{eqnarray}
&& {\cal E}^{(2a)}_f = {\left\langle {\bm u}^2 \right\rangle^{(0)} \over 8 \pi} \,  \meanB_s \, \varepsilon_{fij} \,
\int {E(k) \over k^2}\, \tau^2(k) \, (- {\rm i} k_s)  \, I_{ijmn}
\nonumber\\
&&  \times P_{mn} \, d{\bm k}
= {2 \over 45} \,  \ell_0^2 \,  \meanB_j  \, \lambda_r \,  \Big[3 \varepsilon_{fpm} \Delta_{rmjq} + 2 \varepsilon_{fmr} \, \Delta_{pqmj}
\nonumber\\
&&
+ 5 (\varepsilon_{fjp} \,  \delta_{rq} + \varepsilon_{frp} \,  \delta_{jq}) \Big]   \nabla_p \meanU_{q} ,
\label{D3}
\end{eqnarray}

\begin{eqnarray}
&& {\cal E}^{(2b)}_f = {\left\langle {\bm u}^2 \right\rangle^{(0)} \over 8 \pi}  \meanB_s  \varepsilon_{fij}
\int {E(k) \over k^2} \tau^2(k) (- {\rm i} k_s)  J_{ijmn} P_{mn} \, d{\bm k}
\nonumber\\
&& \quad = {1 \over 45} \,  \ell_0^2 \,  \meanB_j  \, \lambda_r \,
\Big[3 \varepsilon_{fpm} \Delta_{rmjq} + 2 \varepsilon_{fmr} \, \Delta_{pqmj}
\nonumber\\
&& \quad+ 5 (\varepsilon_{fjp} \,  \delta_{rq} + \varepsilon_{frp} \,  \delta_{jq}) \Big] \nabla_p \meanU_{q} ,
\label{D4}
\end{eqnarray}

\begin{eqnarray}
&& {\cal E}^{(2c)}_f = - {\left\langle {\bm u}^2 \right\rangle^{(0)} \over 8 \pi}  \meanB_j  \lambda_r \varepsilon_{fij}
\int {E(k) \over k^2}   \tau^2(k)  I_{irmn} \, P_{mn} \, d{\bm k}
\nonumber\\
&& \quad  = {4 \over 45} \,  \ell_0^2 \,  \meanB_j  \, \lambda_r \Big[
(3+ q)\varepsilon_{fnj} \, \Delta_{pqnr}
- 5 q \, \varepsilon_{frj}\,\delta_{pq}\Big]
\nabla_p \meanU_{q} ,
\nonumber\\
\label{DDD3}
\end{eqnarray}
where $\ell_0^2=\left\langle {\bm u}^2 \right\rangle^{(0)} \, \tau_0^2$,  and
$\Delta_{ijmn} = \delta_{ij} \delta_{mn} + \delta_{im} \delta_{jn} + \delta_{in} \delta_{jm}$.

For the integration over angles in the ${\bm k}$ space, we used the following integrals:
\begin{eqnarray}
&&\int_{0}^{2\pi} \, d\varphi \int_{0}^{\pi} k_{ij} \, \sin \vartheta \,d\vartheta
 = {4 \pi \over 3} \, \delta_{ij} ,
\label{D5}
\end{eqnarray}

\begin{eqnarray}
&&\int_{0}^{2\pi} \, d\varphi \int_{0}^{\pi} k_{ijmn} \, \sin \vartheta \,d\vartheta
= {4 \pi \over 15} \, \Delta_{ijmn} ,
\label{D6}
\end{eqnarray}
where $k_{ijmn} = k_{ij} \, k_{mn}$.
To integrate over $k$, we used the following integral:
$\int_{k_0}^{k_\nu} \tau^2(k) \, E(k) \, dk = 4 \tau_0^2/3$, and
\begin{eqnarray}
&& \int \tau(k) f_{ij}^{(S)}({\bm  k}) \, d{\bm k} = {4 \ell_0^2  \over 45} \Big[
(4q-3) \delta_{ij} {\rm div} \meanUU
\nonumber\\
&& - 2(q+3) (\partial \meanU)_{ij}\Big] .
\label{DBB3}
\end{eqnarray}
We take into account that in anelastic approximation,
$(i k_n - \nabla_n/2) f_{in}({\bm  k}) = - \lambda_n  f_{in}({\bm  k})$.
This implies that the contributions to the turbulent electromotive force caused by the last three terms in equation~(\ref{B3})
is given by
\begin{eqnarray}
\tilde{\cal E}_f &=& \varepsilon_{fij} \overline{B}_j \int \tau(k) \biggl[i k_n - \lambda_n - {1 \over 2} \nabla_n \biggr]f_{in}^{(S)}({\bm  k}) \, d{\bm k}
\nonumber\\
&=& - 2 \lambda_n \varepsilon_{fij} \overline{B}_j \int \tau(k) f_{in}^{(S)}({\bm  k}) \, d{\bm k} = 2 {\cal E}^{(2c)}_f .
\end{eqnarray}

The contributions of the small-scale dynamo in the background turbulence
to the turbulent electromotive force are given by
\begin{eqnarray}
&& {\cal E}^{\rm (M,\lambda)}_f = {\left\langle {\bm b}^2 \right\rangle^{(0)} \over 8 \pi}  \meanB_s  \varepsilon_{fij}
\int {E(k) \over k^2} \tau^2(k) ({\rm i} k_s)  J_{ijmn} P_{mn} d{\bm k}
\nonumber\\
&& \quad = {\ell_0^2 \over 45} \,   \biggl[{\ell_{_{\rm M}} \over \ell_0}\biggr]^{3(q-1)}
\biggl[{ \left\langle {\bm b}^2 \right\rangle^{(0)} \over \mu_0 \meanrho \, \left\langle {\bm u}^2 \right\rangle^{(0)} } \biggr] \,  \meanB_j  \, \lambda_r \,
\Big[3 \varepsilon_{fnp} \Delta_{rnjq}
\nonumber\\
&& \quad - 2 \varepsilon_{fnr} \, \Delta_{pqnj} - 5 (\varepsilon_{fjp} \,  \delta_{rq} + \varepsilon_{frp} \,  \delta_{jq}) \Big] \nabla_p \meanU_{q} ,
\label{MD4}
\end{eqnarray}
\begin{eqnarray}
&& {\cal E}^{(\Lambda_{\rm M})}_f  = -\left\langle {\bm b}^2 \right\rangle^{(0)} \, {\meanB_s \varepsilon_{fij} \over 16 \pi}
\int \tau^2(k) \,  k_s \,  E_{ijmn} \,  E(k) \, k^{-4}
\nonumber\\
&& \times \Big(\Lambda^{\rm (M)}_m k_{n} - \Lambda^{\rm (M)}_n k_{m}\Big) \,  d{\bm k}
= - {2  \ell_0^2\over 45}  \biggl[{\ell_{_{\rm M}} \over \ell_0}\biggr]^{3(q-1)} \, \meanB_j \Lambda^{\rm (M)}_r
\nonumber\\
&&
\times \biggl[{ \left\langle {\bm b}^2 \right\rangle^{(0)} \over \mu_0 \meanrho \, \left\langle {\bm u}^2 \right\rangle^{(0)} } \biggr]
\, \Big[2 (q-1) \varepsilon_{frn} \, \Delta_{pqjn} + 5  \Big(\varepsilon_{fqj} \,  \delta_{rp}
\nonumber\\
&&+\varepsilon_{fjr} \,  \delta_{qp} \Big) \Big] \nabla_p \meanU_{q} ,
\label{MD1}
\end{eqnarray}
where we take into account that $\int_{k_{_{\rm M}}}^{k_\nu} ... E(k) \,dk = \int_{0}^{\tau_{_{\rm M}}} ... \,d\tilde \tau$, $\tau_{{_{\rm M}}} = (\ell_{_{\rm M}} / \ell_0)^{q-1}$, $\tilde \tau(k) =(k/k_0)^{1-q}$, $k_{_{\rm M}}=\ell_{_{\rm M}}^{-1}$ and $\ell_\nu = k_\nu^{-1} \to 0$.

Now we take into account that ${\cal E}_i = a_{ij} \, \meanB_j$, so that
the corresponding contributions to the tensor $a_{ij}$ are given by
\begin{eqnarray}
&& a_{ij}^{(1a)} = {2 \ell_0^2 \over 45} \, \biggl[
10 \left(\tilde {\bm \lambda} \cdot \meanWW\right) \,  \delta_{ij}
-5 (\tilde \lambda_i \, \meanW_j
+ \tilde \lambda_i \, \meanW_j)
\nonumber\\
&& \quad - 2 \tilde \lambda_m \, \Big((4q
-2) \varepsilon_{inm} \, (\partial \meanU)_{nj}
+ (2q-1) \varepsilon_{ijm} \, {\rm div} \meanUU
\nonumber\\
&& \quad
- 5 \varepsilon_{ijn} \, \, (\partial \meanU)_{nm}
\Big) \biggr],
\label{AD1}
\end{eqnarray}

\begin{eqnarray}
&& a_{ij}^{(1b)} = -{4 \ell_0^2 \over 45} \tilde \lambda_m \biggl[(4q+7)
\varepsilon_{inm} \, (\partial \meanU)_{nj}
+ {5 \over 2} \Big[\left(\tilde {\bm \lambda} \cdot \meanWW\right) \,  \delta_{ij}
\nonumber\\
&& \quad
- \tilde \lambda_j \, \meanW_i \Big]
+ (2q+6) \varepsilon_{ijm} \, {\rm div} \meanUU\biggr],
\label{AD2}
\end{eqnarray}

\begin{eqnarray}
&& a_{ij}^{(2a)} =2 a_{ij}^{(2b)} = {2 \ell_0^2 \over 45} \, \biggl[\left({\bm \lambda} \cdot \meanWW\right) \,  \delta_{ij}
+ \lambda_i \, \meanW_j + \lambda_j \, \meanW_i
\nonumber\\
&& \quad + 2 \lambda_m \, \Big(\varepsilon_{inm} \, (\partial \meanU)_{nj} + \varepsilon_{ijn} \, (\partial \meanU)_{mn}
+ \varepsilon_{ijm} \, {\rm div} \meanUU\Big) \biggr],
\nonumber\\
\label{AD4}
\end{eqnarray}

\begin{eqnarray}
&&a_{ij}^{(2c)} = {4 \over 45} \ell_0^2  \, \varepsilon_{ijm} \, \lambda_n \Big[
(4q-3) \, \delta_{mn} \, {\rm div} \meanUU
\nonumber\\
&& \quad - 2(q+3) (\partial \meanU)_{mn}\Big] ,
\label{ADD3}
\end{eqnarray}

The total contribution to the turbulent electromotive force
caused by a density-stratified
homogeneous ($\Lambda_i=0$)
turbulence with a large-scale shear for a low Mach numbers is
\begin{eqnarray}
&& a_{ij}^{(\lambda)}  =  a_{ij}^{(1a)} + a_{ij}^{(1b)} + a_{ij}^{(2a)} + a_{ij}^{(2b)}
+ 2 a_{ij}^{(2c)}
\nonumber\\
&&\quad = {\ell_0^2  \over 45} \biggl[13 \left({\bm \lambda} \cdot \meanWW\right) \,  \delta_{ij} + 3 \, \meanW_i  \, \lambda_j
- 7\meanW_j  \, \lambda_i
\nonumber\\
&&\quad  - 2 \lambda_m \, \Big(
(4 q - 7) \, \varepsilon_{inm} (\partial \meanU)_{nj}
- (14q-22)  \, \varepsilon_{ijm} \, {\rm div} \meanUU
\nonumber\\
&& \quad +
(8q+11) \, \varepsilon_{ijn}  (\partial \meanU)_{mn}\Big) \biggr].
\label{MD1}
\end{eqnarray}
The total contribution to the turbulent electromotive force
caused by an inhomogeneous turbulence with a non-uniform large-scale flow
for a low Mach numbers is ${\cal E}_i^{(\Lambda)}
= a_{ij}^{(\Lambda)} \, \meanB_j$ and $a_{ij}^{(\Lambda)}=a_{ij}^{(1a)}(\lambda=0) + a_{ij}^{(1b)}(\lambda=0)$, i.e.,
\begin{eqnarray}
&& a_{ij}^{(\Lambda)}  =  {\ell_0^2  \over 45} \biggl[5 \, \meanW_j  \, \Lambda_i
-5 \left({\bm \Lambda} \cdot \meanWW\right) \,  \delta_{ij}
\nonumber\\
&&\;
+2 \Lambda_m \, \Big(
(4 q - 2) \, \varepsilon_{inm} (\partial \meanU)_{nj} +
(4q+5) \, \varepsilon_{ijm} \, {\rm div} \meanUU
\nonumber\\
&&\;
- 5 \, \varepsilon_{ijn} \, (\partial \meanU)_{nm} \Big) \biggr] .
\label{MD2}
\end{eqnarray}

The contributions of the small-scale dynamo in the background turbulence
to the tensor $a_{ij}$ are given by
\begin{eqnarray}
&&
a_{ij}^{\rm (M,\lambda)} =- {\ell_0^2 \over 45} \biggl({\ell_{_{\rm M}} \over \ell_0}\biggr)^{3(q-1)}
\biggl[{ \left\langle {\bm b}^2 \right\rangle^{(0)} \over \mu_0 \meanrho \, \left\langle {\bm u}^2 \right\rangle^{(0)} } \biggr]
\biggl[\left({\bm \lambda} \cdot \meanWW\right) \,  \delta_{ij}
\nonumber\\
&& \quad
+ \lambda_i \, \meanW_j + \lambda_j \, \meanW_i + 2 \lambda_m \, \Big(\varepsilon_{inm} \, (\partial \meanU)_{nj}
\nonumber\\
&& \quad
+ \varepsilon_{ijn} \, (\partial \meanU)_{mn} + \varepsilon_{ijm} \, {\rm div} \meanUU\Big) \biggr],
\label{BAD4}
\end{eqnarray}
\begin{eqnarray}
&& a_{ij}^{(\Lambda_{\rm M})} ={2 \ell_0^2 \over 45} \biggl({\ell_{_{\rm M}} \over \ell_0}\biggr)^{3(q-1)}
\biggl[{ \left\langle {\bm b}^2 \right\rangle^{(0)} \over \mu_0 \meanrho \, \left\langle {\bm u}^2 \right\rangle^{(0)} } \biggr] \,
\biggl[ {5 \over 2} \Big(\Lambda^{\rm (M)}_j \, \meanW_i
\nonumber\\
&& \quad
- \Lambda^{\rm (M)}_i \, \meanW_j\Big) + \Lambda^{\rm (M)}_m
\Big(4(q-1) \varepsilon_{inm} \, (\partial \meanU)_{nj}
\nonumber\\
&& \quad
+ 5 \varepsilon_{ijn} \, (\partial \meanU)_{mn} + (2q-7)\varepsilon_{ijm} \, {\rm div} \meanUU\Big) \biggr] .
\label{MBAD4}
\end{eqnarray}

To derive equations~(\ref{AD1})--(\ref{MBAD4}),
we use the following identities:
\begin{eqnarray}
&&\varepsilon_{inp} \Delta_{jnqr} \left(\nabla_p \meanU_{q} \right) \lambda_r =  {1 \over 2} \Big[\left({\bm \lambda} \cdot \meanWW\right) \,  \delta_{ij} + \meanW_j  \, \lambda_i
\nonumber\\
&& \quad  - 4 \meanW_i  \,\lambda_j \Big] + \Big[\varepsilon_{ijn}  (\partial \meanU)_{mn} + \varepsilon_{imn} (\partial \meanU)_{nj} \Big] \lambda_m   ,
\label{D15}
\end{eqnarray}

\begin{eqnarray}
&&\varepsilon_{inr} \Delta_{pqnj} \left(\nabla_p \meanU_{q} \right) \lambda_r = \lambda_m \,
\Big[2 \varepsilon_{inm} (\partial \meanU)_{nj}
\nonumber\\
&& \quad + \varepsilon_{ijm} \, {\rm div} \meanUU \Big] ,
\label{D16}
\end{eqnarray}

\begin{eqnarray}
&&\varepsilon_{ijn} \Delta_{pqnr} \left(\nabla_p \meanU_{q} \right) \lambda_r = \lambda_m \,
\Big[2 \varepsilon_{ijn} (\partial \meanU)_{nm}
\nonumber\\
&& \quad + \varepsilon_{ijm} \, {\rm div} \meanUU \Big] ,
\label{DD16}
\end{eqnarray}

\begin{eqnarray}
&&\varepsilon_{ijp} \left(\nabla_p \meanU_{q} \right) \lambda_q = {1 \over 2} \Big[ \meanW_j \lambda_i  - \meanW_i \lambda_j \Big]
+ \varepsilon_{ijp} (\partial \meanU)_{pq} \lambda_q .
\nonumber\\
\label{D17}
\end{eqnarray}

\begin{eqnarray}
&&\varepsilon_{irp} \left(\nabla_p \meanU_{j} \right) \lambda_r =  {1 \over 2} \Big[\left({\bm \lambda} \cdot \meanWW\right) \,  \delta_{ij} - \meanW_i  \, \lambda_j \Big]
\nonumber\\
&& \quad  + \varepsilon_{imn}  (\partial \meanU)_{nj} \lambda_m   ,
\label{D18}
\end{eqnarray}

\begin{eqnarray}
&&\varepsilon_{iqr} \left(\nabla_j \meanU_{q} \right) \lambda_r =
{1 \over 2} \Big[\left(\meanWW \cdot {\bm \lambda}\right) \,  \delta_{ij} - \meanW_i \lambda_j \Big]
\nonumber\\
&& \quad + \varepsilon_{iqm} (\partial \meanU)_{qj} \lambda_m ,
\label{D19}
\end{eqnarray}

\begin{eqnarray}
&&\varepsilon_{ijq} \left(\nabla_p\meanU_{q} \right) \lambda_p =
{1 \over 2} \Big[ \meanW_i \lambda_j  - \meanW_j \lambda_i \Big]
\nonumber\\
&& \quad + \varepsilon_{ijq} (\partial \meanU)_{pq} \lambda_p .
\label{D20}
\end{eqnarray}

To take into account the compressible contributions (for arbitrary Mach numbers) to the turbulent electromotive force,
we use the following identities:
\begin{eqnarray}
&& {\cal E}^{(3a)}_f = {\left\langle {\bm u}^2 \right\rangle^{(0)} \sigma_c \over 4 \pi (1 + \sigma_c)} \,  \meanB_s \, \varepsilon_{fij}
\int {E(k) \over k^2} \,  \tau^2(k) \, (- {\rm i} k_s) \, I_{ijmn}
\nonumber\\
&& \;  \times  k_{mn}  \,  d{\bm k}= {4 \,\sigma_c\over 45\,(1 + \sigma_c)} \,  \ell_0^2 \,  \meanB_j  \, \lambda_r \,
\varepsilon_{fnp} \Delta_{jnqr} \nabla_p \meanU_{q} ,
\nonumber\\
\label{LR1}
\end{eqnarray}

\begin{eqnarray}
&& {\cal E}^{(3b)}_f = {\left\langle {\bm u}^2 \right\rangle^{(0)} \sigma_c \over 4 \pi (1 + \sigma_c)} \,  \meanB_s \, \varepsilon_{fij}
\int {E(k) \over k^2} \,  \tau^2(k) \, (- {\rm i} k_s)\,  J_{ijmn}
\nonumber\\
&& \times k_{mn}  \,  d{\bm k}= {2 \,\sigma_c\over 45\,(1 + \sigma_c)} \,  \ell_0^2 \,  \meanB_j  \, \lambda_r \,
\varepsilon_{fnp} \Delta_{jnqr} \nabla_p \meanU_{q} ,
\nonumber\\
\label{LR2}
\end{eqnarray}

\begin{eqnarray}
&& {\cal E}^{(3c)}_f = - {\left\langle {\bm u}^2 \right\rangle^{(0)} \sigma_c \over 4 \pi (1 + \sigma_c)} \, \meanB_j \, \lambda_s \varepsilon_{fij} \int {E(k) \over k^2} \tau^2(k) I_{ismn}
\nonumber\\
&& \times k_{mn}  \,  d{\bm k}= {4 (2q+1) \, \sigma_c \over 45\,(1 + \sigma_c)} \,  \ell_0^2 \,  \meanB_j  \, \lambda_s\,  \varepsilon_{fjn} \Delta_{pqns} \, \nabla_p \meanU_{q} ,
\nonumber\\
\label{LR3}
\end{eqnarray}

\begin{eqnarray}
&& {\cal E}^{(3d)}_f = {\sigma_c \left\langle {\bm u}^2 \right\rangle^{(0)} \over 4 \pi (1 + \sigma_c)} \,  \meanB_j \, \varepsilon_{fij}
\int {E(k) \over k^2} \,  \tau^2(k) \, \biggl[{\rm i} k_s - {1 \over 2} \Lambda_s\biggr]
\nonumber\\
&& \quad \times I_{ismn} k_{mn}  \,  d{\bm k}
={2  \sigma_c \, \ell_0^2 \over 45\,(1 + \sigma_c)}  \meanB_j
\varepsilon_{fnj} \biggl[
\lambda_r \,\Big(5 \delta_{np} \delta_{qr}
\nonumber\\
&& \quad - \Delta_{npqr}\Big)  - (2q+1) \Lambda_r \Delta_{npqr} \biggr]\, \nabla_p \meanU_{q} ,
\label{LR11}
\end{eqnarray}
so that
\begin{eqnarray}
&& a_{ij}^{(3a)} = 2 a_{ij}^{(3b)} = {2\ell_0^2 \over 45} \left({\sigma_c  \over 1 + \sigma_c}\right) \biggl[\left({\bm \lambda} \cdot \meanWW\right) \,  \delta_{ij} + \lambda_i \, \meanW_j
\nonumber\\
&& \; - 4\lambda_j \, \meanW_i + 2 \lambda_m \, \Big(\varepsilon_{imn} \, (\partial \meanU)_{nj} + \varepsilon_{ijn} \, (\partial \meanU)_{mn}
\Big)\biggr],
\nonumber\\
\label{LR4}
\end{eqnarray}

\begin{eqnarray}
&& a_{ij}^{(3c)} = {4 (2q+1) \over 45} \, \ell_0^2 \, \left({\sigma_c  \over 1 + \sigma_c}\right)  \, \varepsilon_{ijn}
\Big[2\lambda_m \, (\partial \meanU)_{mn}
\nonumber\\
&& \;+ \lambda_n \, \, {\rm div} \meanUU \Big] .
\label{LR5}
\end{eqnarray}

\begin{eqnarray}
&& a_{ij}^{(3d)} = - {2 \ell_0^2 \over 45} \left({\sigma_c  \over 1 + \sigma_c}\right) \biggl[
{5 \over 2} \Big( \lambda_i \, \meanW_j - \lambda_j \, \meanW_i \Big)
\nonumber\\
&& \quad
-  \lambda_m\varepsilon_{ijm} \, {\rm div} \meanUU + 3 \lambda_m \varepsilon_{ijn} (\partial \meanU)_{mn}
\nonumber\\
&& \quad
-  (2 q + 1) \, \varepsilon_{ijn} \Big(2\Lambda_m \, (\partial \meanU)_{nj}
+\Lambda_n \, {\rm div} \meanUU \Big)\biggr] .
\label{LR10}
\end{eqnarray}
Therefore,
\begin{eqnarray}
a_{ij}^{\rm (tot)} &=& a_{ij}(\sigma_c=0) + a_{ij}^{(\sigma_c)} ,
\label{MD3}
\end{eqnarray}
where
\begin{eqnarray}
a_{ij}(\sigma_c=0) &=&  a_{ij}^{(1a)} + a_{ij}^{(1b)} + a_{ij}^{(2a)} + a_{ij}^{(2b)}
+ 2 a_{ij}^{(2c)} ,
\label{LR7}
\end{eqnarray}
and
\begin{eqnarray}
&& a_{ij}^{(\sigma_c)} = - {\sigma_c  \over 1 + \sigma_c} \,\Big(a_{ij}^{(2a)} + a_{ij}^{(2b)} + 2a_{ij}^{(2c)}\Big)
+ a_{ij}^{(3a)}
\nonumber\\
&&  \quad + a_{ij}^{(3b)} + a_{ij}^{(3c)} + a_{ij}^{(3d)} = - {\ell_0^2 \over 45} \left({\sigma_c  \over 1 + \sigma_c}\right)
\biggl[5 \lambda_i \, \meanW_j
\nonumber\\
&&  \quad  + 10 \lambda_j \, \meanW_i
+ 2 \lambda_m \Big[2 (2 q -3)\varepsilon_{ijm} \, {\rm div} \meanUU
\nonumber\\
&& \quad + 6 \varepsilon_{inm} (\partial \meanU)_{nj} - (12 q +13) \varepsilon_{ijn} (\partial \meanU)_{mn}\Big]
\nonumber\\
&&  \quad -  2 (2 q + 1) \, \varepsilon_{ijn} \Big(2\Lambda_m \, (\partial \meanU)_{nm}
+\Lambda_n \, {\rm div} \meanUU \Big)\biggr] .
\label{CMD3}
\end{eqnarray}

\end{document}